\documentclass[nobibnotes,aps,superscriptaddress,10pt,notitlepage,longbibliography,nofootinbib]{revtex4-1}
\usepackage[utf8]{inputenc}
\usepackage[T1]{fontenc}
\usepackage{comment}
\usepackage{graphicx}
\usepackage{amssymb}
\usepackage{amsmath}
\usepackage{amsfonts}
\usepackage{physics}
\usepackage[colorlinks=true,citecolor=blue,hyperfootnotes=false]{hyperref}
\usepackage[dvipsnames]{xcolor}
\usepackage{dcolumn}
\usepackage{textcomp}
\usepackage{xcolor}
\usepackage{xfrac}
\usepackage{slashed}
\usepackage{multirow}
\usepackage[scr=esstix]{mathalpha}
\usepackage{float}

\def\be{\begin{equation}}
\def\ee{\end{equation}}
\def\beq{\begin{equation}}
\def\eeq{\end{equation}}
\def\figs/B{B}
\def\bea{\begin{eqnarray}}
\def\eea{\end{eqnarray}}
\def\bg{\begin{eqnarray}}
\def\nd{\end{eqnarray}}

\def\sin{{\rm sin}}
\def\cos{{\rm cos}}

\def\ln{{\rm log}}

\def\be{\begin{equation}}
\def\ee{\end{equation}}

\def\sin{{\rm sin}}
\def\cos{{\rm cos}}

\def\ln{{\rm ln}}

\def\dd{{\rm d}}

\makeatletter
\renewcommand{\thesubsection}{\thesection~\Alph{subsection}}
\def\p@subsection{}

\let\oldaddcontentsline\addcontentsline
\renewcommand{\addcontentsline}[3]{%
  \begingroup
  \def\thesubsection{\Alph{subsection}}%
  \oldaddcontentsline{#1}{#2}{#3}%
  \endgroup
}
\makeatother

\begin{document}
% --------------------------------%
%%%%%%%%%%%%%%%%%%%%%%%%%%%%%%%%%%%

\hfill RIKEN-iTHEMS-Report-26

\vskip 0.5in

\title{Interacting Dark Energy and Dark Matter in O(3) No-Scale Gravity }

\author{Lincoln da S. Pereira}
\affiliation{Department of Mathematical Physics,
Institute of Physics, University of São Paulo, R. do Matão 1371, São Paulo, SP 05508-090, Brazil}
\affiliation{Kavli Institute for the Physics and Mathematics of the Universe (WPI),
UTIAS, The University of Tokyo, Chiba 277-8583, Japan}

\author{Muzi Hong}
\affiliation{RIKEN Center for Interdisciplinary Theoretical and Mathematical Sciences (iTHEMS), 
RIKEN, Wako, 351-0198, Japan}

\author{Elisa G. M. Ferreira}
\affiliation{Kavli Institute for the Physics and Mathematics of the Universe (WPI),
UTIAS, The University of Tokyo, Chiba 277-8583, Japan}

\author{Tsutomu T. Yanagida}
\affiliation{Kavli Institute for the Physics and Mathematics of the Universe (WPI),
UTIAS, The University of Tokyo, Chiba 277-8583, Japan}
\affiliation{Tsung-Dao Lee Institute \& School of Physics and Astronomy, Shanghai Jiao Tong University, Pudong New Area, Shanghai 201210, China}

\begin{abstract}
Dark Energy (DE) and Dark Matter (DM) are among the greatest mysteries in particle physics and cosmology, since their origins remain unclear. It is intriguing to consider whether both may originate from a purely gravitational sector. We propose that they arise from degrees of freedom associated with the partners of the Brans–Dicke boson in No-Scale Gravity, a fundamental scale-invariant gravitational theory in which the Planck scale is illusional. We consider the Brans--Dicke boson, together with two scalar fields, to form a vector multiplet under an $O(3)$ symmetry, whose angular directions in the Einstein frame are identified with DM and DE. Small explicit breaking of $O(3)$ generates their masses and interaction, and we show that the resulting model reproduces the required background cosmological evolution and present-day abundances, with the lighter field providing a dynamical DE component at late times. Although the DE field remains canonical, the interacting cosmology exhibits an effective phantom-like evolution at late times, of the type suggested by DESI in combination with other cosmological probes, without introducing a fundamental phantom degree of freedom. The model provides a first-principles realization of interacting DE, in which DM, DE, and their coupling emerge from the symmetry structure of the underlying gravitational theory.
\end{abstract}

\maketitle

\tableofcontents

%%%%%%%%%%%%%%%%%%%%%%%%%%%%%%%%%%%
%%%%%%%%%%%%%%%%%%%%%%%%%%%%%%%%%%%
%%%%%%%%%%%%%%%%%%%%%%%%%%%%%%%%%%%
%%%%%%%%%%%%%%%%%%%%%%%%%%%%%%%%%%%
\section{Introduction}\label{sec:intro}
%%%%%%%%%%%%%%%%%%%%%%%%%%%%%%%%%%%
%%%%%%%%%%%%%%%%%%%%%%%%%%%%%%%%%%%
%%%%%%%%%%%%%%%%%%%%%%%%%%%%%%%%%%%
%%%%%%%%%%%%%%%%%%%%%%%%%%%%%%%%%%%

Dark Matter (DM) and Dark Energy (DE) are among the greatest mysteries in particle physics and cosmology. A pressureless cold fluid description of DM is remarkably successful in reproducing large-scale observations, but its fundamental origin remains unknown.  A cosmological constant as the DE candidate provides the simplest explanation for the late-time accelerated expansion.  Recent results from the Dark Energy Spectroscopic Instrument (DESI)~\cite{DESI:2013agm,DESI:2016fyo,DESI:2016igz,DESI:2025zpo,DESI:2025zgx}, particularly when combined with other cosmological probes, have provided tentative indications that DE may evolve with time rather than behave as a strict cosmological constant.
Many models of dynamical DE have been proposed, several of which provide a good description of the evolution preferred by current combinations of DESI, CMB, and supernova data~\cite{DESI:2013agm,DESI:2016fyo,DESI:2016igz,DESI:2025zpo,DESI:2025zgx}. In these models, the evolution of DE is generally determined by the dynamics of an additional field. Another possibility is that the evolution of DE is affected by an interaction with DM. Interacting dark-sector models can also provide a good description of current cosmological observations~\cite{Giare:2024smz,Antusch:2026ldp,Guedezounme:2025wav}. One of their attractive features is that the interaction can give rise to an effective phantom behaviour, with equation of state (EoS) $w_{\rm eff}<-1$, even when the DE field itself does not cross the phantom divide~\cite{Das:2005yj}. A phantom-like expansion history can therefore be obtained without introducing a fundamental phantom field.

Despite this success, many dynamical-DE and interacting dark-sector models remain phenomenological, with the dark components and, when present, their coupling introduced separately. Given the phenomenological success of these models, it is important to develop frameworks in which DM, DE, and their interaction emerge from a common underlying theory. In this paper, we explore this possibility within \textit{No-Scale Gravity}.

The Planck scale $M_{p}$ is the most important and fundamental mass parameter in Einstein gravity. No-Scale Gravity \cite{Hong:2025tyi} is a theory without any dimensionful parameters, and the Planck scale is not fundamental but illusional. Due to the absence of any dimensionful parameters, including the cutoff scale at the quantum level, the theory is completely scale invariant \cite{Wetterich:2019qzx}. After the Weyl transformation to the Einstein frame, the theory takes the form of Einstein gravity with a massless boson called the dilaton $\chi$. The massless dilaton couples derivatively to matter, and hence it does not generate the long-range force usually associated with Brans--Dicke theories. 

No-Scale Gravity can therefore be viewed as an extension of Brans--Dicke gravity~\cite{Brans:1961sx} that evades the usual long-range-force constraints~\cite{Will:2005va}. In the Einstein frame, No-Scale Gravity closely resembles Einstein gravity, apart from the presence of the dilaton $\chi$, although its fundamental philosophy is completely different. This motivates us to consider No-Scale Gravity as a fundamental framework.

This makes No-Scale Gravity a natural framework in which the dark sector may arise from degrees of freedom associated with gravity itself. A first realization of this idea was recently developed for dynamical DE by extending the Brans--Dicke scalar to an $O(2)$ multiplet~\cite{Hong:2025cae}. In the Einstein frame, the radial direction is associated with the dilaton, while the angular direction provides a light scalar field that can play the role of DE. In this work, we extend this construction to $O(3)$. The Brans--Dicke field is promoted to a three-component vector, introducing a second angular degree of freedom. This opens the possibility of accommodating both DM and DE within the same framework.

To realize this idea, small explicit breakings of the $O(3)$ symmetry are introduced. These terms generate the masses of the two angular fields and induce an interaction between them, so that DM, DE, and their interaction arise from the same gravitational sector. Different patterns of symmetry breaking lead to different realizations of the model. In this work, we focus on the regime in which the interaction between the two dark components is cosmologically relevant.

Within this framework, we show that the $O(3)$ construction provides a well-motivated interacting dark-sector model with the required cosmological behaviour of DM and dynamical DE. The interaction can also lead to an effective phantom regime, potentially accommodating the evolution suggested by current cosmological data. In particular, we focus on an ultralight DM candidate, which adds further phenomenological interest to the construction. We further investigate how this evolution depends on the model parameters and initial conditions.

This paper is organized as follows. In Section~\ref{sec:model}, we briefly review the No-Scale Gravity model, and present its extension to incorporate interacting DM and DE, as sketched above. In Section~\ref{sec:Cosmology}, we numerically solve the cosmological evolution of DM and DE, and present the EoS and abundances. We also discuss the appearance of a phantom-like behavior in the effective EoS of DE. We conclude the paper in section~\ref{sec:conclusion}.

%%%%%%%%%%%%%%%%%%%%%%%%%%%%%%%%%%%
%%%%%%%%%%%%%%%%%%%%%%%%%%%%%%%%%%%
%%%%%%%%%%%%%%%%%%%%%%%%%%%%%%%%%%%
%%%%%%%%%%%%%%%%%%%%%%%%%%%%%%%%%%%
\section{No-Scale Gravity }\label{sec:model}
%%%%%%%%%%%%%%%%%%%%%%%%%%%%%%%%%%%
%%%%%%%%%%%%%%%%%%%%%%%%%%%%%%%%%%%
%%%%%%%%%%%%%%%%%%%%%%%%%%%%%%%%%%%
%%%%%%%%%%%%%%%%%%%%%%%%%%%%%%%%%%%

The No-Scale Gravity
 \cite{Hong:2025tyi} revolves around the idea that the Planck scale, and all scales, are an illusion. All dimensionful parameters, even the cutoff scale, are replaced by the BD scalar field $\phi$ or by appropriate powers of $\phi$. This makes it a scale-transformation invariant theory \cite{Wetterich:2019qzx}. The Lagrangian in Jordan frame \footnote{Unlike Ref.~\cite{Hong:2025cae}, we do not assume $R^2$ inflation, and therefore $\xi$ is not fixed here by inflationary considerations. Values $\xi=\mathcal{O}(1)$ remain possible and may be relevant for explaining cosmic birefringence through an operator of the form $(\phi_1\phi_2/\phi^2)F\widetilde{F}$~\cite{Lin:2025gne}. In the cosmological analysis below, we instead focus on smaller values of $\xi$, corresponding to the trans-Planckian field-space scale used in our dark-sector benchmarks.} is
\begin{align}
    \mathcal{L} = \frac{\xi}{2}\phi^2 R + \frac{1}{2}\partial_\mu \phi \partial^\mu \phi - \frac{\lambda}{4}\phi^4 + \mathcal{L}_{\rm SM}.
\end{align}
Here, $\xi$  is a positive constant and $\mathcal{L}_{\rm SM}$ is the Standard Model (SM) Lagrangian in which all mass parameters are replaced by $\phi$~\cite{Ferreira:2016kxi, Burrage:2018dvt}. Unlike theories which have a dimensionful cutoff scale, we are able to keep scale invariance at the quantum level~\footnote{This prescription has been introduced as \textit{scale-invariant renormalization scheme} in previous literature~\cite{Englert:1976ep, Shaposhnikov:2008xi,Armillis:2013wya,Hamada:2016onh,Falls:2018olk}.} since the cutoff itself is proportional to $\phi$. 

By taking the Weyl transformation $g_{\mu\nu}^\text{E} = \Omega^2 g_{\mu\nu}$ with $\Omega^2 = \xi \phi^2/M^2_{\rm pl}$ \footnote{
This frame transformation should be distinguished from the global scale transformation in the Jordan frame. The latter is the transformation used to discuss global scale-invariance symmetry in the Jordan frame. The BD field $\phi$ remains unchanged under the former, but transforms under the latter.
}, the action now can be written in the Einstein frame as
\begin{align}
    \mathcal{L} = \frac{M_p^2}{2} R+ \frac{1}{2}M_p^2\qty(6+\frac{1}{\xi}) \partial_\mu \ln{\Omega}\partial^\mu \ln{\Omega}  
    - \frac{\lambda}{4\xi^2} M_p^4
    + \Omega^{-4}\mathcal{L}_{\rm SM},
\end{align}
where we dropped the superscript E for simplicity. The $\lambda\phi^4$ term in the Jordan frame becomes a cosmological constant in the Einstein frame, and we take $\lambda\ll1$.
We can tell from this action that in the Einstein frame, we have a massless field
\begin{equation}
    \chi \equiv M_p \sqrt{6+\frac{1}{\xi}} \rm ln\Omega.
\end{equation}
The scale-invariant symmetry in the Jordan frame becomes the shift symmetry of $\chi$ in the Einstein frame. As for the matter sector, classically, the gauge and fermionic kinetic terms in the SM Lagrangian are Weyl invariant. However, the kinetic term and mass term of the Higgs field are non-Weyl invariant, which requires attention. 
As mentioned above, we replace the Higgs mass term by $\sim \lambda_{h}\phi^2 |H|^2$ in the Jordan frame, which is Weyl invariant with canonical normalization of the Higgs field $H\to \Omega H$.
The long-range force is then not induced, since Higgs now only couples to the massless field $\chi$ through derivative couplings, \textit{e.g.}, $\sim \partial|H|^2\partial \chi$.
Also, since the shift symmetry of $\chi$ is maintained at the quantum level, we do not have the long-range force arising from operators induced by quantum effects after the Weyl transformation (see \cite{Hong:2025tyi} for more details).

\subsection{The $O(3)$ No-Scale Gravity}

The $O(3)$ No-Scale Gravity model is a natural extension of the construction developed in Ref.~\cite{Hong:2025cae}. We generalize the BD scalar to a vector $\Vec{\phi}$ in the fundamental representation of $O(3)$, thereby introducing the additional degrees of freedom required to accommodate both DE and DM. The Lagrangian in the Jordan frame is
    \begin{align}
            {\mathcal{L}} =& \frac{\xi}{2}\phi_i\phi_i R +  \frac{1}{2}\partial_\mu \phi_i \partial^\mu\phi_i - \frac{\lambda (\phi_i\phi_i)^2}{4}   + \mathcal{L}_{\rm SM}, 
            \label{eq:o3BD}
    \end{align}
where $\phi_{1,2,3}$ are three scalar fields and repeated indices are summed over $i=1,2,3$. Motivated by the $O(3)$ symmetry, we introduce polar coordinates,
$\phi_1 = \rho \sin\theta \sin\varphi$, $\phi_2 = \rho \sin\theta \cos\varphi $ and $\phi_3 = \rho \cos\theta$.
The radial direction $\rho$ plays the role of the BD field $\phi$ in the Jordan model.
It is related to the dilaton $\chi$ in the Einstein frame as we will see below in Eq.~(\ref{eq:Lagrangian_polar}).
The two angular directions are also flat in the Einstein frame due to the $O(3)$ symmetry, which will be identified with DE and DM.
We therefore have three massless bosons, including the dilaton $\chi$ in the Einstein frame, 
and we have a symmetric reason why we have two light bosons associated to DM and DE.
Also, from now on we assume the contribution from the cosmological constant is negligible compared with the dynamical DE.

Now we are at the point to introduce the explicit $O(3)$ breaking terms. In order to realize the mass hierarchy, $m_{\rm DM} \gg m_{\rm DE}$, we consider the sequential breaking, $O(3)\to O(2)\to D_4$. Here, $\phi_1$ and $\phi_2$ transforms in the fundamental representation of $O(2)$, and the $D_4$ includes, for example, a parity ($\phi_1\to -\phi_1$ or $\phi_2\to -\phi_2$) and an exchange symmetry between $\phi_1$ and $\phi_2$. Then, we have 4 possible dimension-4 operators which are invariant under $D_4$, these are
\begin{align}
    \epsilon_1(\phi_1^2+\phi_2^2+\phi_3^2)\phi^2_3 \ ;  \qquad \epsilon_2\phi_1^2\phi_2^2 \ ; \qquad \epsilon_3\phi_3^4 \ ; \qquad \epsilon_4(\phi_1^4+\phi_2^4).
    \label{eq:brkterms1}
\end{align}
In polar coordinates, the above breaking terms are expressed as 
\begin{align}
    \epsilon_1\rho^4\cos^2\theta \ ;  \qquad \epsilon_2\rho^4\sin^4\theta \cos^2\varphi\sin^2\varphi \ ; \qquad \epsilon_3\rho^4\cos^4{\theta} \ ; \qquad \epsilon_4\rho^4\sin^4\theta(\sin^4\varphi+\cos^4\varphi).
\end{align}
We see that the $\epsilon_1$ and $\epsilon_3$ breaking terms only generate a potential in the $\theta$ direction. We choose the first term since it has the classical mass term for $\theta$ at the potential minimum. 
As for the second breaking, we choose the second term for the same reason for $\varphi$, and we consider $\epsilon_1 \gg \epsilon_2> 0$. The other terms are ignored in this paper. In this sequential breaking case, the dynamics of the DM and DE are simple since their mutual interaction is small. The above hierarchy suggests a naturally small interaction. 

To make the interaction more relevant, we consider the inverted hierarchy. This is achieved by taking the $\epsilon_2$ term to dominate over the $\epsilon_1$ term. 
In this breaking pattern, the dynamics becomes complicated but interesting, since the mutual interaction between the DM and the DE is large. We will discuss the physics of the DM and the DE in separate subsections below for each scenario, respectively.

\subsection{Sequential hierarchy $O(3)\to O(2)\to D_4$: the weakly interacting limit}

We discuss the physics in the sequential breaking case of $O(3)\to O(2)\to D_4$. 
The starting Lagrangian in the Jordan frame is Eq.~(\ref{eq:o3BD}) and terms with $\epsilon_1 \gg \epsilon_2$ in Eq.~(\ref{eq:brkterms1}), while $\epsilon_3=\epsilon_4=0$.
After the Weyl transformation $g_{\mu\nu}^\text{E} = \Omega^2 g_{\mu\nu}$ with $\Omega^2 = \xi \rho^2/M^2_{\rm pl}$,
the Lagrangian in the Einstein frame is
\begin{align}\label{eq:Lagrangian_polar}
        \mathcal{L} =&\ \frac{M_p^2}{2} R + \frac{1}{2}M_p^2\qty(6+\frac{1}{\xi})\partial_\mu \ln \Omega \partial^\mu \ln \Omega  +\frac{M^2_p}{2\xi}[\partial_\mu \theta \partial^\mu\theta+\sin^2\theta\partial_\mu \varphi \partial^\mu\varphi] \nonumber\\
    &\ - \frac{\lambda M_p^{4}}{4\xi^2} -  \frac{\epsilon_1 M_p^4}{\xi^2}\cos^2\theta - \frac{\epsilon_2 M_p^4}{\xi^2} \sin^4\theta \cos^2\varphi\sin^2\varphi,
    \end{align}
where we dropped the superscript E for simplicity and omitted the SM Lagrangian. To simplify the Lagrangian, we define the following variables:
\begin{align}
    m_\theta^2 \equiv \epsilon_1\frac{M^2_p}{\xi}, \qquad m^2_\varphi \equiv 2\epsilon_2\frac{M^2_p}{\xi}, \qquad f_\theta \equiv \frac{M_p}{\sqrt{\xi}}.
\end{align}
Also, we perform field redefinitions for canonical normalization:

   \begin{align}
    \theta \to \frac{\sqrt{\xi}}{M_p}\theta,\qquad \varphi \to\frac{\sqrt{\xi}}{M_p}\varphi, \qquad \chi \equiv     M_p\sqrt{6+\frac{1}{\xi}} \ln \Omega.
   \end{align}
The Lagrangian can then be expressed in terms of the dilaton $\chi$ and the DM and the DE sector $(\theta,\varphi)$. We also use the identity $\sin^2\theta = [1-\cos(2\theta)]/2$ and $\cos^2\theta = [1+\cos(2\theta)]/2$ to rewrite the potential:  

    \begin{align}\label{eq:Lagrangian_canon}
             \mathcal{L} 
     &=\frac{M_p^2}{2} R + \frac{1}{2}(\partial \chi)^2  +\frac{1}{2}\qty[\partial_\mu \theta \partial^\mu\theta+\sin^2\qty(\frac{\theta}{f_\theta})\partial_\mu \varphi \partial^\mu\varphi]\nonumber \\
    & \ \ \ \ - \frac{\lambda M_p^4}{4\xi^2} -  \frac{m^2_{\theta} f_\theta^2}{2}\qty[1+\cos{\qty(\frac{2\theta}{f_\theta})}]- \frac{m^2_{\varphi} f_\theta^2}{16} \sin^4\qty(\frac{\theta}{f_\theta}) \qty[1-\cos{\qty(4\frac{\varphi}{f_\theta})}].
    \end{align}

As already stated, the cascade breaking induces the mass hierarchy $m_\theta\gg m_\varphi>0$, which leads to $\theta$ and $\varphi$ being identified with DM and DE, respectively. Notice that the minimum of the potential of the DM (\textit{i.e.} $\theta$) is at $\theta/f_\theta={\pi}/2$. Remarkably, the DE (\textit{i.e.} $\varphi$) has a non-vanishing kinetic term and the maximum mass at that DM potential minimum. This is a key point for this model to work in a consistent manner.

The dynamics provided by this setup is simple, since the parameter $\epsilon_2$ in the interaction is orders of magnitude smaller than $\epsilon_1$ in the $m_\theta^2$ terms.
Through most of the universe's history, one expects $\theta$ to behave as DM without interaction with DE. 
Also, since the minimum of the DM potential is $\theta/f_\theta = \pi/2$, one expects that close to today the $\sin(\theta/f_\theta)$ suppression on DE is alleviated, as $\theta$ oscillates around this minimum. 
Hence, for this hierarchical breaking, the model mimics an almost non-interacting axion-like Fuzzy DM + quintessence DE through most of the universe's history. There is though a growth in the DE mass after $\theta$ starts to quickly oscillate, as in this case $m_{\rm DE}^2\sim m_\varphi^2 [1-(\delta\theta_{\rm amp}/f_\theta)^2/2]$, where $\delta\theta_{\rm amp}$ is the amplitude of oscillation~\footnote{This is derived by taking time average of ${\rm sin}^2(\pi/2+\delta\theta/f_\theta)={\rm cos}^2(\delta\theta/f_\theta)$ as $\langle {\rm cos}^2(\delta\theta/f_\theta)\rangle \approx 1-\langle (\delta\theta/f_\theta)^2 \rangle \approx 1-(\delta\theta_{\rm amp}/f_\theta)^2/2$.}. This effect is small since the oscillation amplitude of DM is suppressed at late times due to the expansion of the universe.

Although this hierarchy provides a consistent weakly interacting realization of DM and DE, we do not investigate its cosmological evolution further in this work.

\subsection{Inverse hierarchy: the interacting dark sector}
\label{sec:inv}

In the previous breaking pattern, the interaction term proportional to $\epsilon_2$ is negligible compared to $\epsilon_1$, and the interaction between DM and DE has little impact.
To have significant effects from the interaction, we take the same Lagrangian as Eq.~(\ref{eq:Lagrangian_polar}) but with the inverted hierarchy $\epsilon_1\ll\epsilon_2$ ($m_\theta\ll m_\varphi$) for which the interaction between the two angular fields can have significant cosmological effects. The heavier field $\varphi$ is identified with DM, while the lighter field $\theta$ is identified with DE.

\begin{figure}[t]
        \centering
        \includegraphics[width=1.0\linewidth]{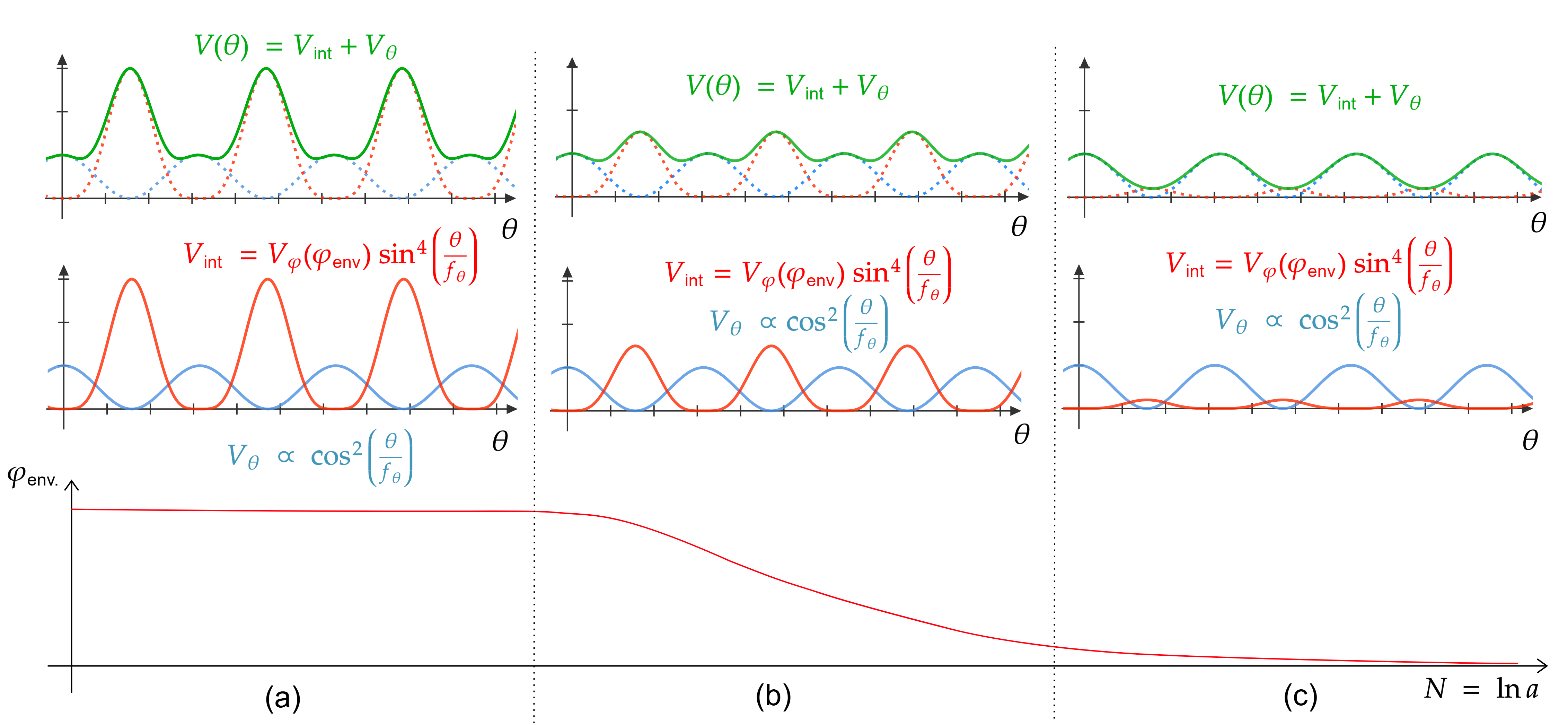}
        \caption{
    Potential as a function of $\theta$ 
    with different field values or oscillation amplitudes of $\varphi$.
    The plots are divided into 3 regions from left to right: (a) $\varphi$ is frozen; (b) $\varphi$ starts oscillating; and (c) $\varphi$ oscillates near its potential minimum.
    On the bottom, we have the plot of the $\varphi$ envelope evolution ($\varphi_{\rm env.}$) in terms of e-folds.
    In the middle row, we show $V_\theta$ in blue and $V_{\rm int}$ in red, which are defined in Eq.~(\ref{eq:potential}). 
    In the top panels, we show the sum of these two contributions as $V(\theta)$ in green.
    }
        \label{fig:Interaction_pot}
\end{figure}

To discuss the effect of this change in the hierarchy of the masses, we separate the potential into two parts.
We define $V(\varphi,\theta)=V_\theta+V_{\rm int}$, with
    \begin{align}\label{eq:potential}
        V_\theta \equiv   \frac{\lambda M_p^4}{4\xi^2} + \frac{m^2_{\theta} f_\theta^2}{2}\qty[1+\cos{\qty(\frac{2\theta}{f_\theta})}], \qquad V_\varphi \equiv  \frac{m^2_{\varphi} f_\theta^2}{16} \qty[1-\cos{\qty(4\frac{\varphi}{f_\theta})}], \qquad  V_{\rm int} \equiv \sin^4\qty(\frac{\theta}{f_\theta})V_\varphi.
    \end{align}
The coefficient in $V_{\rm int}$ is now large, and we cannot ignore the interaction.
The contribution to the potential of $\theta$ is first dominated by the interaction term if the initial angle |$\varphi_i/f_\theta$| is not too small. After the oscillation amplitude of $\varphi$ decayed, $V_{\rm int}$ is suppressed by the $\varphi$ contribution, and the contribution to the potential of $\theta$ is now dominated by $V_\theta$. This change of the potential of $\theta$ with time evolution is shown in Figure \ref{fig:Interaction_pot}.
At late times, $\theta$ rolls toward its potential minimum $\pi/2$.
However, in the early time, the minimum is near $\theta=0$, which is a coordinate singularity due to the fields living in $\mathbb{S}^2$, which nullify $\varphi$ completely. This is an issue of the chart choice, which does not fully cover the field space and is not an actual physical singularity. 
If the initial value of DE is not properly taken care of, it might hit the singularity before its potential is dominated by $V_\theta$.

This interacting realization is the model that we study throughout the remainder of this work. We restrict the analysis to the region in which the chart is well defined and take the initial value of the DE field close to $\pi/2$. In the following section, we study its cosmological evolution by numerically solving the coupled field equations.

%%%%%%%%%%%%%%%%%%%%%%%%%%%%%%%%%%%
%%%%%%%%%%%%%%%%%%%%%%%%%%%%%%%%%%%
%%%%%%%%%%%%%%%%%%%%%%%%%%%%%%%%%%%
%%%%%%%%%%%%%%%%%%%%%%%%%%%%%%%%%%%
\section{Background cosmology with the inverse hierarchy breaking}\label{sec:Cosmology}
%%%%%%%%%%%%%%%%%%%%%%%%%%%%%%%%%%%
%%%%%%%%%%%%%%%%%%%%%%%%%%%%%%%%%%%
%%%%%%%%%%%%%%%%%%%%%%%%%%%%%%%%%%%
%%%%%%%%%%%%%%%%%%%%%%%%%%%%%%%%%%%

% \subsection{DE and DM cosmological evolution }
\subsection{Cosmological equations}
%%1. intro

 We now study the cosmological evolution of the interacting dark-sector model introduced in the previous section. The two angular fields $\varphi$ and $\theta$ play the roles of DM and DE, respectively. In the Einstein frame, the gravitational sector takes the standard Einstein--Hilbert form, while the dynamics of the dark sector are governed by their coupled two-field Lagrangian. We assume a spatially flat Friedmann--Lemaître--Robertson--Walker spacetime and take both scalar fields to be homogeneous. The metric is therefore given by
\begin{equation}
    ds^2 = dt^2-a^2(t)\left(dx^2+dy^2+dz^2\right),
\end{equation}
where $a(t)$ is the scale factor and $t$ denotes proper time.

% 2. EoM in proper time
We take $\theta=\theta(t)$ and $\varphi=\varphi(t)$, and denote derivatives with respect to $t$ by overdots. The equations of motion are
\begin{align}
    &\Ddot{\theta} + 3H\dot{\theta}-\frac{1}{2f_{\theta}}\sin\qty(2\frac{\theta}{f_{\theta}}) \dot{\varphi}^2- m^2_{\theta}f_\theta\sin\qty(2\frac{\theta}{f_{\theta}})+\frac{m^2_{\varphi}f_\theta}{4}\qty[1-\cos\qty(4\frac{\varphi}{f_{\theta}})]\sin^3\qty(\frac{\theta}{f_{\theta}})\cos\qty(\frac{\theta}{f_{\theta}}) = 0\label{eq:prop_theta}\,,\\
    &\Ddot{\varphi} + 3H\dot{\varphi}+\frac{2}{f_{\theta}}\cot\qty(\frac{\theta}{f_{\theta}}) \dot{\varphi}\dot{\theta} + \frac{m_{\varphi}^2f_\theta}{4 }\sin\qty(4\frac{\varphi}{f_{\theta}})\sin^2\qty(\frac{\theta}{f_{\theta}}) = 0\label{eq:prop_phi}\,.
\end{align}
Here, $H$ is the Hubble parameter.
In addition to the usual Hubble-friction terms, the equations contain kinetic couplings arising from the nontrivial geometry of the two-dimensional field space. The terms proportional to $\dot{\varphi}^{2}$ and $\dot{\theta}\dot{\varphi}$ therefore couple the evolution of the two angular fields, even independently of their potential interaction. The potential $V_{\rm int}$ introduces a further coupling.

The evolution of $\theta$ receives contributions from both its own potential and its interaction with $\varphi$, with their relative importance changing over time. At sufficiently early times, when the Hubble rate is larger than the characteristic frequencies of the scalar fields, Hubble friction overdamps their motion and keeps them approximately frozen. As the Universe expands and the Hubble rate decreases, $\theta$ begins to evolve under the competition between these two contributions.
%%%% 3. energy densities, EoS, ...

To characterize the two dark components, we follow the decomposition of the potential introduced in Eq.~(\ref{eq:potential}). Since the two components interact, they are not separately conserved, and the assignment of the interaction energy between them is not unique. We choose to include $V_{\rm int}$ in the DM component. We therefore associate the kinetic energy and potential $V_\theta$ of $\theta$ with DE, and the field-dependent kinetic energy of $\varphi$, together with $V_{\rm int}$, with DM. We define
    \begin{align}
        \rho_\theta &= \frac{1}{2} \dot{\theta}^2+\frac{m^2_\theta f_\theta^2}{2}\qty[1+\cos \qty(\frac{2\theta}{f_\theta})]+\frac{\lambda M_p^4}{4\xi^2}\,,\label{eq:rho_th}\\
        \rho_\varphi &= \frac{1}{2}\sin^2\qty(\frac{\theta}{f_\theta})\dot{\varphi}^2   + \frac{m^2_{\varphi} f_\theta^2}{16} \sin^4\qty(\frac{\theta}{f_\theta}) \qty[1-\cos \qty(4\frac{\varphi}{f_\theta})]\,,
        \label{eq:rho_ph}
    \end{align}
with respective pressure
\begin{align}
    P_\theta &= \frac{1}{2} \dot{\theta}^2-\frac{m^2_\theta f_\theta^2}{2}\qty[1+\cos \qty(\frac{2\theta}{f_\theta})]-\frac{\lambda M_p^4}{4\xi^2}\,, \label{eq:wtheta}\\
    P_\varphi &=   \frac{1}{2}\sin^2\qty(\frac{\theta}{f_\theta})\dot{\varphi}^2   - \frac{m^2_{\varphi} f_\theta^2}{16} \sin^4\qty(\frac{\theta}{f_\theta}) \qty[1-\cos \qty(4\frac{\varphi}{f_\theta})]\,.
\end{align}
With these definitions, the equations of state of the two components are $w_\theta=P_\theta/\rho_\theta$ and $w_\varphi=P_\varphi/\rho_\varphi$. Together with the baryon and radiation components, these quantities determine the background expansion through the Friedmann equations.

As the expressions above illustrate, the energy density of DM depends on the value of $\sin(\theta/f_\theta)$, which can seemingly render $\rho_\varphi=0$ if $\sin(\theta/f_\theta)=0$. This is, of course, not a physical problem, but a consequence of our coordinate chart not being defined there. This does not affect our numerical simulations, as the singular point is never reached during the evolution considered here.

%%%% 4. Eq. in e-folds
For the numerical evolution, it is convenient to use the number of e-folds, $N=\ln a$, as the time variable. We set $a=1$ today, so that $N=0$ at the present time, and denote derivatives with respect to $N$ by primes. The subscript zero denotes the present values. This yields the following equations of motion:  

    \begin{widetext}
    \begin{align}
        &\theta'' + \qty(3+\dv{\ln H}{N})\theta'-\frac{1}{2f_\theta}\sin\qty(2\frac{\theta}{f_\theta}) \varphi'^2 -\frac{m^2_{\theta}f_\theta}{H^2}\sin\qty(2\frac{\theta}{f_\theta})+\frac{m^2_{\varphi}f_\theta}{4H^2}\qty[1-\cos\qty(4\frac{\varphi}{f_\theta})]\sin^3\qty(\frac{\theta}{f_\theta})\cos\qty(\frac{\theta}{f_\theta})  = 0,\\
        &\varphi'' + \qty(3+\dv{\ln H}{N})\varphi'+\frac{2}{f_\theta}\cot\qty(\frac{\theta}{f_\theta}) \varphi'\theta' + \frac{m_{\varphi}^2f_{\theta}}{4H^2 }\sin\qty(4\frac{\varphi}{f_\theta})\sin^2\qty(\frac{\theta}{f_\theta}) = 0.\label{eq:phi_eom}
    \end{align}
    \end{widetext}
The background expansion is determined by the Friedmann equations, which can be written in terms of the fields and their derivatives as
    \begin{widetext}
    \begin{align}
        &  H^2 = \frac{ \frac{m^2_\theta f_\theta^2 }{2}\qty[1+\cos \qty(2\frac{\theta}{f_\theta})] + \frac{\lambda M_{\rm pl}^4}{4\xi^2}  + \frac{m^2_{\varphi} f_\theta^2 }{16} \sin^4\qty(\frac{\theta}{f_\theta}) \qty[1-\cos \qty(4\frac{\varphi}{f_\theta})] + 3M^2_{p}\Omega_b^0 H_0^2 e^{-3N}+ 3M^2_{p}\Omega_r^0 H_0^2 e^{-4N}}{3M^2_{p}-\frac{1}{2}\qty[ \theta'^2+\sin^2\qty(\frac{\theta}{f_\theta})\varphi'^2] },\\
        &  \dv{\ln H}{N} =-\frac{1}{2M_{p}^2}\qty[ \theta'^2+\sin^2\qty(\frac{\theta}{f_\theta})\varphi'^2] -\frac{1}{2H^2}\qty( 3\Omega_b^0 H_0^2 e^{-3N}+ 4\Omega_r^0 H_0^2 e^{-4N}).
    \end{align}
    \end{widetext}

%%%% 5. Rapid oscil
The cosmological roles of the two fields are controlled by the mass hierarchy entering the system above. In our construction, $m_\varphi \gg m_\theta$. To realize dynamical DE, we take $m_\theta \lesssim H_0$, so that the evolution of $\theta$ is overdamped by Hubble friction and the field remains approximately frozen during most of the cosmological history. It begins to evolve only at late times, when the Hubble rate becomes comparable to its characteristic mass scale.

The DM field $\varphi$ is heavier. In this work, we focus on the ultralight regime, with masses around $m_\varphi\sim10^{-20}\,\mathrm{eV}$, sometimes called fuzzy DM. This particular scale is not fixed by the construction, but depends on the chosen symmetry-breaking parameter. We chose here to be ultralight DM (as we will discuss in the next section). Such an ultralight field $\varphi$, at early times, is also overdamped by Hubble friction and remains approximately frozen. As the Hubble rate decreases, the field eventually begins to oscillate coherently around the minimum of its potential.  For the axion-like potential considered here, the potential is approximately quadratic near the minimum, and the oscillating field has an average EoS $\langle w_\varphi\rangle\simeq0$, behaving as pressureless DM~\cite{Marsh:2015xka, Ferreira:2020fam, Hui:2021tkt, Eberhardt:2025caq}.

For an ultralight DM mass of order $m_\varphi\sim10^{-20}\,\mathrm{eV}$, the oscillation frequency soon becomes much larger than the Hubble rate. Resolving each oscillation throughout the cosmological evolution is therefore numerically inefficient. We instead use the standard envelope approximation, which follows the slowly varying amplitude of the field while averaging over its rapid oscillations.

To coarse-grain the quick oscillations, we take the envelope approximation where we choose the ansatz to be $\varphi \sim \varphi_{\rm amp} \cos\int^N\dd N' \omega(N')$. 
Expanding Eq.~\eqref{eq:phi_eom} around the potential minimum,
$\varphi=0$, where $\sin(4\varphi/f_\theta)\simeq4\varphi/f_\theta$, we identify the
instantaneous oscillation frequency $\omega$ and effective friction $\Gamma$ as $\omega^2 = (m^2_\varphi/H^2)\sin^2(\theta/f_\theta)$ and $\Gamma= 3+H'/H+2{\rm cot}(\theta/f_\theta)\theta'/f_\theta$, respectively.
 Since the oscillation is fast, we have $|\varphi'_{\rm amp}/\varphi_{\rm amp}| \ll\omega$, $|\omega'/\omega| \ll\omega$ and $|\Gamma|\ll \omega$.
With the quickly oscillating $\varphi$, we have $\varphi'_{\rm amp}+(\omega'/\omega+\Gamma)\varphi_{\rm amp}/2\approx0$. 
For the approximation of $\varphi$ in the equations of $\theta$ and the Hubble constant, we take a time average over the quickly oscillating $\varphi$, \textit{i.e.}, $\langle \varphi'^2\rangle\approx \omega^2\varphi^2_{\rm amp}/2$ and $\langle \varphi^2\rangle\approx \varphi^2_{\rm amp}/2$.

Using these oscillation averages, together with $\omega'/\omega=-H'/H+\cot(\theta/f_\theta)\theta'/f_\theta$, the equations for $\theta$ and the slowly varying amplitude become
 \begin{align}
    &\theta'' + \qty(3+\dv{\ln H}{N})\theta'- \frac{m^2_{\theta}f_\theta}{H^2}\sin\qty(2\frac{\theta}{f_\theta})+\frac{m^2_{\varphi}}{2H^2 f_\theta}\varphi_{\rm amp}^2\sin^3\qty(\frac{\theta}{f_\theta})\cos\qty(\frac{\theta}{f_\theta}) = 0\,,\label{eq:appr1}\\
    &\varphi_{\rm amp}'+\frac{3}{2}\qty[1 +\frac{1}{f_\theta}\cot\qty(\frac{\theta}{f_\theta})\theta']\varphi_{\rm amp}=0 \label{eq:amp}\,.
\end{align}
The averaged Friedmann equations read:
\begin{align}
    & H^2 = \frac{ \frac{m^2_\theta f_\theta^2 }{2}\qty[1+\cos \qty(2\frac{\theta}{f_\theta})] + \frac{\lambda M_p^4}{4\xi^2}   + 3M^{2}_{p}\Omega_b^0 H_0^2 e^{-3N}+ 3M^{2}_{p}\Omega_r^0 H_0^2 e^{-4N}+ \frac{m^2_{\varphi} }{2}\varphi_{\rm amp}^2 \sin^4\qty(\frac{\theta}{f_\theta})}{3M^{2}_{p}-\frac{1}{2} \theta'^2 }\,,\label{eq:appr2}\\
    &  \dv{\ln H}{N} =-\frac{1}{2M^2_{p}} \theta'^2 -\frac{1}{2H^2}\qty[\frac{m^2_\varphi }{2M^2_{p}}\varphi_{\rm amp}^2\sin^4\qty(\frac{\theta}{f_\theta})+ 3\Omega_b^0 H_0^2 e^{-3N}+ 4\Omega_r^0H_0^2e^{-4N}]\,.\label{eq:appr3}
\end{align}
Here, we use a matching condition for the solution before and after the approximation.
The timing of switching to the coarse-grained equations is when $\varphi$ starts to oscillate quickly.
In practice, we take $m_\varphi|\sin(\theta/f_\theta)|=\alpha H$, corresponding to $\omega=\alpha$, with $\alpha\simeq100$. We impose $\varphi_{\rm amp}=|\varphi|$ at the last peak before this condition is met.

In this approximation, the averaged energy densities of the two scalar components are
\begin{align}
        \expval{\rho_\theta} &= \frac{1}{2} \dot{\theta}^2+\frac{m^2_\theta f_\theta^2}{2}\qty[1+\cos \qty(\frac{2\theta}{f_\theta})]+\frac{\lambda M_p^4}{4\xi^2},\label{eq:rho_th_avg}\\
        \expval{\rho_\varphi} &= \frac{1}{2}m_\varphi^2 \varphi_{\rm amp}^2\sin^4\qty(\frac{\theta}{f_\theta}).\label{eq:rho_ph_avg}
\end{align}
After averaging, the expression for $w_\theta$ is unchanged, while the coarse-grained EoS of $\varphi$ becomes $w_\varphi=0$, since its averaged kinetic and potential energy densities are equal. Nevertheless, $\langle\rho_\varphi\rangle$ retains a dependence on $\theta$ through the factor $\sin^4(\theta/f_\theta)$. The DM energy density, therefore, does not scale exactly as $a^{-3}$, unlike that of a separately conserved pressureless component. As can be seen from Eq.~\eqref{eq:amp}, the deviation is controlled by the evolution of $\theta$ and becomes negligible when $\theta'$ is suppressed. This is present in the behavior of $\varphi_{\rm amp}$, however, there is also an overall $\sin^4(\theta/f_\theta)$ suppression as can be seen in Eq. (\ref{eq:rho_ph_avg}).

%%%% 6. IC

From the expression for the energy densities, one can estimate the initial field values by matching to today's abundances. These estimates provide starting values for the numerical determination of the initial conditions. We therefore require $\rho_\theta \sim 3M_{p}^2H_0^2 \Omega_{\rm DE}^{0}$ and $\rho_\varphi a^{3}_{\rm osc} \sim 3M_{p}^2H_0^2 \Omega_{\rm DM}^{0} $ at initial time, where $a_{\rm osc}$ is the scale factor at the moment $\varphi$ starts to oscillate ($m_\varphi \left|\sin (\theta / f_\theta)\right| \sim 3H$), and we normalize $a_0=1$. 
Since the energy densities are dominated by their potential contributions at early times, we can estimate the initial field values as:
    \begin{align}
            \frac{\theta_i}{f_\theta} &\sim \frac{1}{2}\arccos\qty[\frac{6H_0^2\Omega_{\rm DE}^{0}}{m^2_\theta}\qty(\frac{M_{p}^2}{f^2_{\theta}})-1-\frac{\lambda f_\theta^2}{2m_\theta^2}]
            , \label{eq:init_th}\\
            \frac{\varphi_i}{f_\theta} &\sim \sqrt{\frac{6H_0^2\Omega_{\rm DM}^0 }{m^2_{\varphi}\sin^4\qty(\frac{\theta_i}{f_\theta})a^{3}_{\rm osc}}}\frac{M_{p}}{f_{\theta}}.\label{eq:init_ph}
    \end{align}
Since these relations neglect the subsequent interaction between the two fields, they should be understood as approximate estimates. The initial conditions used in the numerical evolution are chosen so that the desired present-day DM and DE abundances are reproduced.

To realize the present cosmic acceleration, the DE field needs to satisfy $m_\theta \lesssim H_0$, otherwise, it oscillates too soon.
Assuming the cosmological constant is negligible, for Eq.~(\ref{eq:init_th}) to have a solution, one then needs $\frac{M_p^2}{f_\theta^2} \lesssim \frac{m_\theta^2}{3H_0^2\Omega_{\rm DE}^0} \lesssim
\frac{1}{3\Omega_{\rm DE}^0},$ where the second inequality follows from $m_\theta\lesssim H_0$.
 This suggests a decay constant near or larger than the Planck scale
\footnote{ 
A non-negligible cosmological constant leads to other possibilities of the value of $f_\theta$, which we will not consider in this paper.}.
We therefore consider a trans-Planckian decay constant in the numerical simulation. A trans-Planckian $f_\theta$ is naturally accommodated in No-Scale Gravity and is related to the small nonminimal coupling required for successful Starobinsky inflation~\cite{Hong:2025tyi}, although no specific inflationary realization is assumed in this work. However, if the ultralight DM field is present during high-scale inflation, the realization of both Starobinsky inflation and fuzzy DM at the same time can lead to the isocurvature problem, since $H_{\rm inflation}$ is large.
One may instead consider relatively low-energy inflation, such as a hill-top inflation (\textit{e.g.}, \cite{Kawasaki:2023zpd}), where the Hubble constant during inflation is small enough to avoid the isocurvature problem for the Fuzzy DM.

With these initial conditions and the matching prescription described above, we now solve the cosmological equations numerically and examine the resulting evolution of the two dark-sector fields.

\subsection{Numerical results and discussion}\label{sec:Numerics}

% 1. Benchmark parameters and ICs used
We now solve the cosmological equations numerically and study the evolution of the interacting dark sector. We first consider a representative solution with $m_\varphi=10^{13}H_0$, $m_\theta=2\times10^{-1}H_0$, and $\xi=10^{-3}$, where $H_0\sim10^{-33}\,\mathrm{eV}$, corresponding to $m_\varphi\sim10^{-20}\,\mathrm{eV}$. The analytical estimates derived above are used to guide the choice of the initial field values, which are taken to be $\theta_i/f_\theta\simeq\pi/2-0.2$ and $\varphi_i/f_\theta\simeq6\times10^{-4}$ and chosen to reproduce the present-day DM and DE abundances given by \cite{Planck:2018vyg}. We initialize the evolution at
$N_i=-40$  ($z_i\sim 2.3\cdot10^{17}$) in the radiation-dominated era, with $\theta_i'=\varphi_i'=0$, since both fields are initially overdamped by Hubble friction. For the transition to the coarse-grained evolution, we use the matching prescription described above with $\alpha=100$, a number that has been validated in the literature~\cite{Hlozek:2014lca,Poulin:2018dzj}.

\begin{figure}[H]
    \centering
    \includegraphics[width=1.0\linewidth]{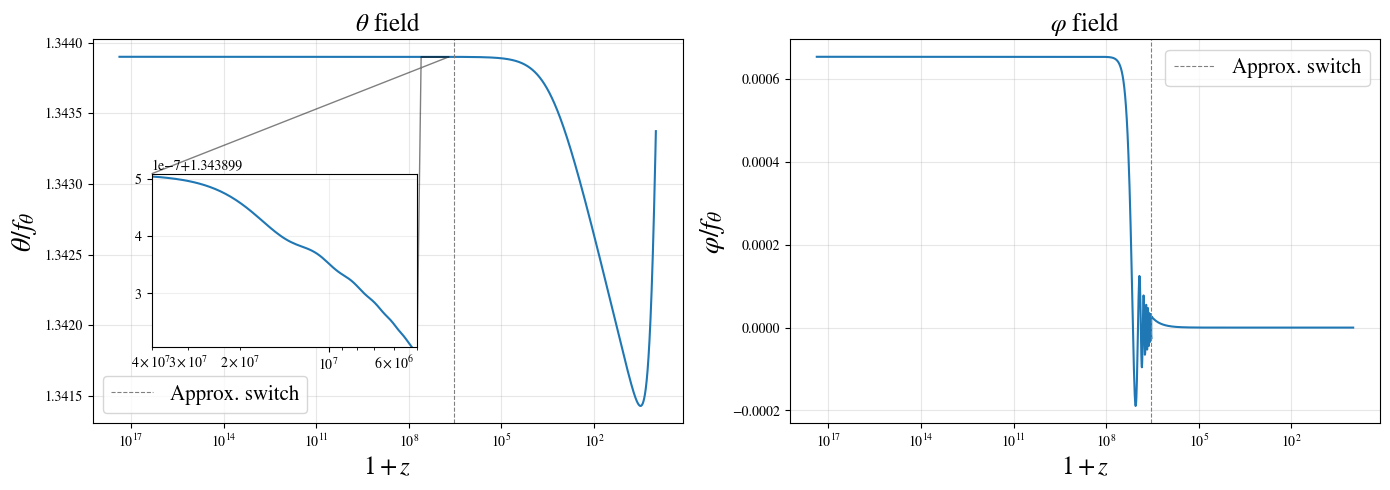}
    \caption{On the left panel, the field value evolution for $\theta$, the DE candidate, as a function of the redshift is plotted, and on the right panel, the field value evolution for $\varphi$, the DM candidate, is plotted. The parameters are taken as $m_\varphi = 10^{13}H_0$, $m_\theta=2\times 10^{-1}H_0$ with $H_0 \sim10^{-33}\,$eV, $\xi = 10^{-3}$ and initial conditions are taken as $(\theta_i/f_\theta;\varphi_i/f_\theta) \sim (\frac{\pi}{2}-0.2; 6\cdot10^{-4})$. The vertical dashed line denotes the transition between the exact solution and the envelope approximation.}
    \label{fig:Field_value_FuzzyDM}
\end{figure}

\begin{figure}[H]
    \centering
    \begin{minipage}{0.48\textwidth}
        \centering
        \includegraphics[width=\textwidth]{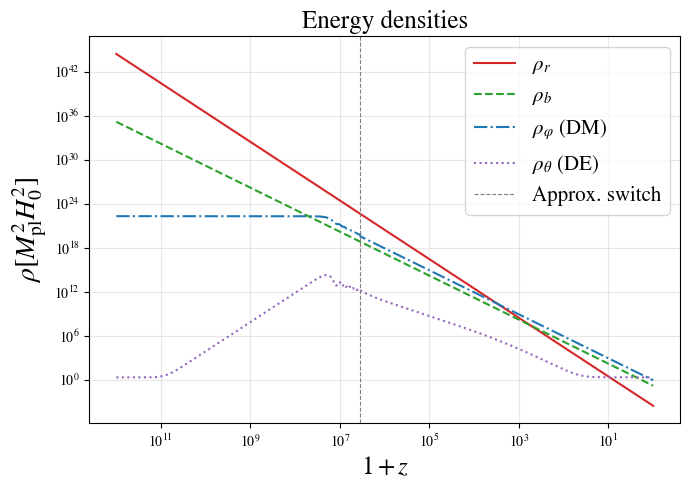}
    \end{minipage}
    \hfill 
    \begin{minipage}{0.48\textwidth}
        \centering
        \includegraphics[width=\textwidth]{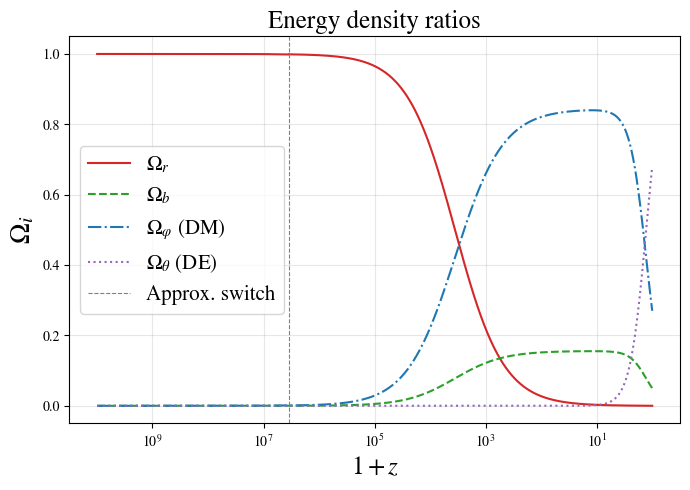}
    \end{minipage}
        \caption{On the left panel, the energy densities as functions of the redshift are plotted, and on the right panel, the abundances are plotted.
        The parameters are taken as 
        $m_\varphi = 10^{13}H_0$, $m_\theta=2\times 10^{-1}H_0$ with $H_0 \sim10^{-33}\,$eV, $\xi = 10^{-3}$ and initial conditions are taken as $(\theta_i/f_\theta;\varphi_i/f_\theta) \sim (\frac{\pi}{2}-0.2; 6\cdot10^{-4})$. The vertical dashed line denotes the transition between the exact solution and the envelope approximation. }
    \label{fig:rho+ratios_fuzzyDM}
\end{figure}

\begin{figure}[H]
    \centering
    \includegraphics[width=1.0\linewidth]{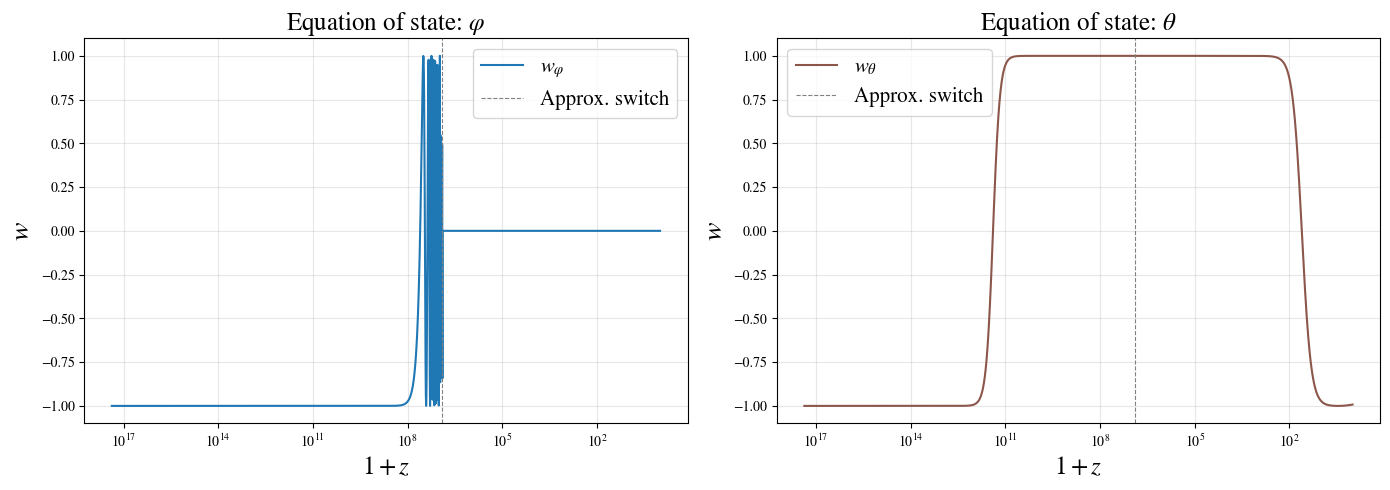}
    \caption{On the left panel, the EoS $w_\varphi$ for the DM candidate as a function of the redshift is plotted, and on the right panel, the EoS for $w_\theta$ for the DE candidate is plotted. The parameters are taken as 
       $m_\varphi = 10^{13}H_0$, $m_\theta=2\times 10^{-1}H_0$ with $H_0 \sim10^{-33}\,$eV, $\xi = 10^{-3}$ and initial conditions are taken as $(\theta_i/f_\theta;\varphi_i/f_\theta) \sim (\frac{\pi}{2}-0.2; 6\cdot10^{-4})$. The vertical dashed line denotes the transition between the exact solution and the envelope approximation. }
    \label{fig:EoS_FuzzyDM}
\end{figure}

Figures~\ref{fig:Field_value_FuzzyDM}, \ref{fig:rho+ratios_fuzzyDM} and \ref{fig:EoS_FuzzyDM} show, respectively, the evolution of the field values, the energy densities and abundances, and the equations of state. The vertical dashed line marks the transition from the exact solution to the envelope approximation.

%%% 2. Evolution of the fields
We first consider the evolution of the two fields, shown in Figure~\ref{fig:Field_value_FuzzyDM}. Initially, both fields are overdamped by Hubble friction. The DM field $\varphi$ remains nearly frozen until its effective oscillation frequency becomes comparable to the Hubble rate. It then
begins to oscillate coherently around the minimum of its potential, realizing the scalar-field DM behaviour with EoS $\langle w_\varphi\rangle\simeq0$.

The DE field $\theta$ is initialized close to $\theta/f_\theta=\pi/2$ and is initially almost frozen by Hubble friction. The interaction with $\varphi$, however, is already present. Although $\theta/f_\theta=\pi/2$ is a minimum of the self-potential $V_\theta$, it is a maximum of the interaction potential for a fixed nonzero value of $\varphi$. At early times, the interaction dominates the curvature of the total effective potential, making this point unstable. Hubble friction strongly suppresses the motion, but $\theta$ nevertheless begins to move slowly away from its initial position, even before $\varphi$ enters the rapid-oscillation regime and before the transition to the envelope approximation.

Once $\varphi$ begins to oscillate, its decreasing amplitude gradually suppresses the interaction contribution. The curvature of the total effective potential around $\theta/f_\theta=\pi/2$ therefore changes sign: the point evolves from an unstable maximum into a stable minimum as the self-potential $V_\theta$ becomes dominant. This change in the effective potential drives $\theta$ back toward $\pi/2$. Since $m_\theta$ is of order $H_0$, this evolution becomes appreciable only at late times, when $\theta$ begins its slow dynamical evolution and the onset of an oscillation, around the minimum, giving rise to the dynamical DE behaviour.

Once $\varphi$ begins to oscillate, its oscillations periodically modulate the interaction force and induce small oscillations in $\dot{\theta}$. Consequently, $\theta$ moves periodically faster and slower, as can be seen more clearly in the inset of the left panel of Figure~\ref{fig:Field_value_FuzzyDM}. These rapid features are no longer resolved after the transition to the envelope approximation, which retains only their averaged effect.

%%% 3. Evolution of the densities, ...
We now turn to the evolution of the energy densities, abundances, and equations of state, shown in Figures~\ref{fig:rho+ratios_fuzzyDM} and~\ref{fig:EoS_FuzzyDM}. At early times, the Universe is radiation-dominated, while both scalar-field components remain subdominant. Before the onset of its oscillations, the energy density of $\varphi$ is approximately frozen. Once $\varphi$ begins to oscillate coherently, its energy density redshifts approximately as $a^{-3}$ and its EoS oscillates around the average value $\langle w_\varphi\rangle\simeq0$. The $\varphi$ component, therefore, behaves as DM and gives rise to the matter-dominated epoch.

The energy density of $\theta$ remains subdominant throughout most of the cosmological evolution. At late times, when the self-potential $V_\theta$ becomes dominant, $\theta$ evolves, with the mass $m_\theta$ setting the timescale of its motion and subsequent oscillations around the minimum. Its EoS therefore departs from that of a cosmological constant, and we then have a dynamical DE component. The energy density of $\theta$ eventually becomes dominant and drives the late-time accelerated expansion. The numerical solution thus reproduces the standard sequence of radiation, matter, and DE domination, together with the observed present-day abundances.

Interestingly, the transition to this late-time evolution is linked to the earlier dynamics of the DM field. As the oscillation amplitude of $\varphi$ decreases, the interaction contribution to the effective potential of $\theta$ is suppressed, allowing its self-potential to become dominant. Although the interaction becomes negligible at late times, its early-time effect determines the displacement and subsequent evolution of $\theta$. The timing of this transition, therefore, depends on the initial field values and on the model parameters. The model provides a dynamical connection between the DM and DE sectors, although it does not by itself remove the sensitivity associated with the cosmic coincidence problem.

In addition to this overall cosmological evolution, the interaction between $\theta$ and $\varphi$ produces several nontrivial transient features. The most prominent is a bump in the DE energy density, accompanied by a period during which the EoS of $\theta$ approaches the stiff value $w_\theta\simeq1$, as we can see in the right panel of Figure~\ref{fig:EoS_FuzzyDM}.

During this period, the interaction term $V_{\rm int}$ dominates over the self-potential $V_\theta$ and drives the motion of $\theta$. The resulting kinetic energy becomes sufficiently large to dominate both $\rho_\theta$ and $p_\theta$. Indeed, when $\dot{\theta}^{2}/2\gg V_\theta$, one has $\rho_\theta = p_\theta \simeq\frac{1}{2}\dot{\theta}^{2}$ and therefore $w_\theta\simeq1$. The enhancement of the kinetic energy is also responsible for the transient bump in $\rho_\theta$ seen in Figure~\ref{fig:rho+ratios_fuzzyDM}.

Despite this stiff behaviour, the energy density of $\theta$ remains orders of magnitude smaller than that of the dominant cosmological components during this period. For the benchmark solution, we find $\left.\rho_\theta/\rho_r\right|_{\rm BBN}\sim10^{-16}$, well below the corresponding BBN bound~\cite{Dutta:2010cu}. This transient behaviour, therefore, has no observable effect on the background expansion. We find the same qualitative result for all the parameter choices explored here: the contribution remains extremely subdominant, as expected for a component that only becomes cosmologically relevant at late times. At late times, $V_\theta$ becomes dominant over the interaction contribution, and $\theta$ returns to a quintessence-like regime.

To quantify the early evolution leading to this transient, we expand the equation of motion around the initial position by writing
\begin{equation}
\theta=\frac{\pi f_\theta}{2}-\Delta\theta,
\qquad
\Delta\theta\ll f_\theta.
\end{equation}
While the interaction dominates over the self-potential, the equation for the displacement becomes approximately
\begin{align}
\ddot{\Delta\theta}
+3H\dot{\Delta\theta}
-\frac{m_\varphi^2}{4}
\left[
1-\cos\left(4\frac{\varphi}{f_\theta}\right)
\right]
\Delta\theta
\simeq0.
\label{eq:theta_approx}
\end{align}
The negative sign of the last term shows that the interaction gives $\Delta\theta$ a tachyonic effective mass squared while $\varphi$ is frozen at its nonzero initial value. The displacement therefore grows as the Hubble friction decreases, eventually generating the kinetic-energy-dominated phase described above. In the overdamped regime, the onset of the transient stiff phase can then be estimated analytically. Neglecting $\ddot{\Delta\theta}$ in Eq.~\eqref{eq:theta_approx} gives $|\dot{\theta}|\sim [m_\varphi^2/(12H)]\left[1-\cos(4\varphi/f_\theta)\right]\Delta\theta$. Although $\theta$ is still overdamped, its velocity therefore increases as the Hubble rate decreases. The stiff phase begins when its kinetic energy becomes comparable to its self-potential energy. For $\varphi/f_\theta\ll1$, this gives, up to factors of order unity, $H_{\rm stiff}\sim(m_\varphi^2/m_\theta)(\varphi_i/f_\theta)^2$. For the benchmark parameters, $H_{\rm stiff}\sim10^{-7}m_\varphi^2/m_\theta\sim10^7m_\varphi$, consistent with the numerical solution and confirming that the transient begins before $\varphi$ enters the rapid-oscillation regime.

\vspace{0.3cm}

\paragraph*{\textbf{Effective phantom behaviour}}

\begin{figure}[H]
    \centering
    \begin{minipage}{0.48\textwidth}
        \centering
        \includegraphics[width=\textwidth]{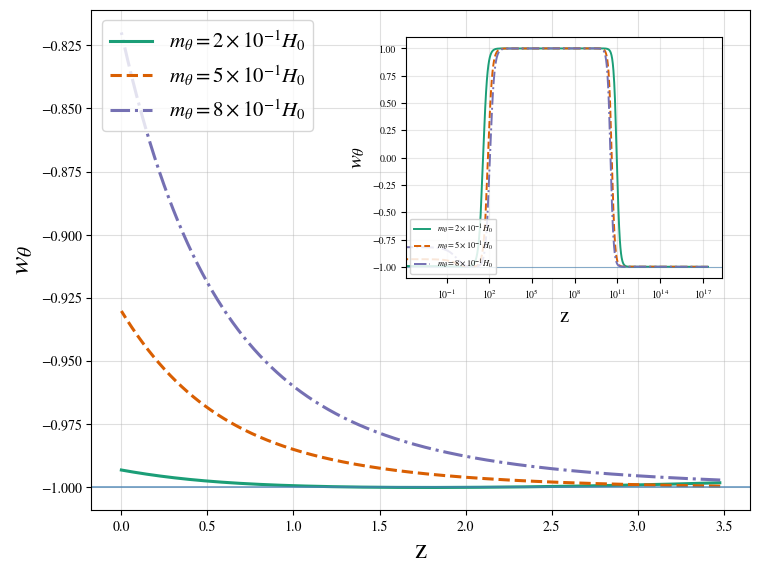}
    \end{minipage}
    \hfill 
    \begin{minipage}{0.48\textwidth}
        \centering
        \includegraphics[width=\textwidth]{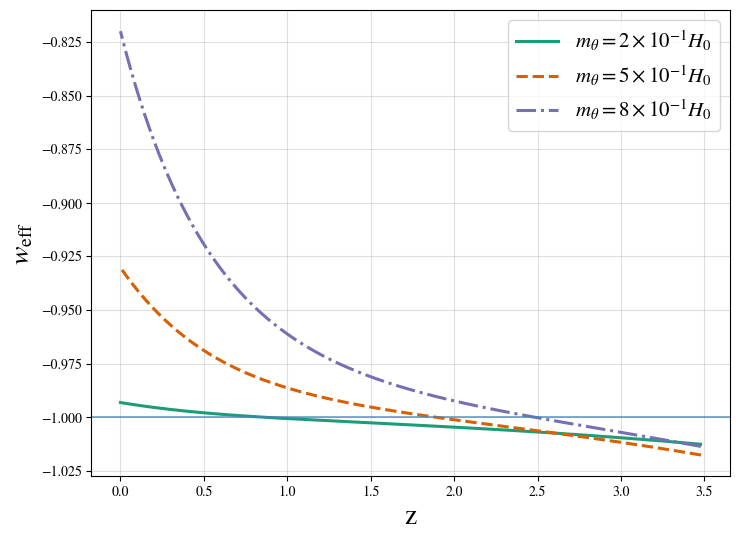}
    \end{minipage}
        \caption{ 
        The EoS and the effective EoS for $\theta$ for the masses $m_\theta \in\{2 , 5,8\}\times 10^{-1}H_0$, with $m_\varphi = 10^{13}H_0$ and $\xi = 10^{-3}$. All initial conditions were chosen to match today's abundances. The left panel shows the EoS of $\theta$ as defined in Eq.~(\ref{eq:wtheta}), for redshift $z<3.5$, with a subplot showing the full evolution of $w_\theta$. The right panel shows the plot for the effective EoS $w_{\rm eff}^{\theta}$ as defined in Eq.~(\ref{eq:effective}). 
        The effective EoS shows phantom behavior, while $w_\theta$ never features $w<-1$. }
    \label{fig:eos_comp}
\end{figure}

We now show that the DE behaviour in this construction can also resemble the phantom-like evolution suggested by current DESI results. As shown in the left panel of Figure~\ref{fig:eos_comp}, the rolling of $\theta$ leads to an evolving EoS of DE. However, $w_\theta$ never crosses into the phantom region, as expected for a canonical scalar field with a positive kinetic term.

An effective phantom behaviour can nevertheless arise if the cosmological evolution is interpreted under the standard assumption that DM is a non-interacting pressureless fluid, with energy density scaling as $a^{-3}$. In this description, the difference between the actual interacting DM density and the density of a separately conserved DM component is absorbed into an effective DE density, defined as $\rho_{\rm DE}^{\rm eff}=\rho_\theta+\rho_\varphi-\rho_{\varphi,0}a^{-3}$. The corresponding effective EoS is then given by~\cite{Das:2005yj}
    \begin{align}\label{eq:effective}
        w_{\rm eff }^{\theta} = \frac{w_\theta}{1 + \qty[\rho_\varphi - \rho_\varphi^0a^{-3}]\frac{1}{\rho_\theta}}\,.
    \end{align}

The effective phantom behaviour shown in the right panel of Figure~\ref{fig:eos_comp} originates from the $\sin^4(\theta/f_\theta)$ suppression of the DM energy density in Eq.~\eqref{eq:rho_ph}. As shown in Figure~\ref{fig:Field_value_FuzzyDM}, $\theta$ rolls towards $\pi f_\theta/2$ at late times, gradually lifting this suppression. Before the present epoch, one therefore has $\rho_\varphi<\rho_{\varphi,0}a^{-3}$, with $\rho_\varphi$ approaching $\rho_{\varphi,0}a^{-3}$ at late times. The correction in the denominator of Eq.~\eqref{eq:effective} is consequently negative, allowing $w_{\rm eff}^{\theta}$ to cross below $-1$ even though the physical EoS always satisfies $w_\theta\geq-1$.

Figure~\ref{fig:eos_comp} also shows that the time of the effective phantom crossing depends on $m_\theta$, which controls the late-time evolution of $\theta$ and, consequently, the suppression of the DM energy density. Varying $m_\theta$ mainly changes the duration and redshift range of the effective phantom phase, while the depth of the phantom excursion remains approximately unchanged for the values considered. For the range of masses considered, the crossing occurs approximately within $0.5 \lesssim z\lesssim 2.5$.

The effective EoS is introduced only to characterize the late-time evolution in an equivalent description in which DM and DE are assumed to be non-interacting. Therefore, the apparent phantom crossing does not correspond to a fundamental phantom degree of freedom and does not indicate any physical pathology. 

Such an apparent phantom crossing without a fundamental phantom degree of freedom is a known feature of interacting dark-sector models, in which the EoS inferred under the assumption of separately conserved DM can differ from the intrinsic DE EoS. Interacting models have also been shown to reproduce the type of effective phantom evolution favoured by DESI and to provide competitive fits to the cosmological data~\cite{Giare:2024smz,Antusch:2026ldp,Guedezounme:2025wav}. Our construction provides a fundamental realization of this phenomenology within No-Scale Gravity, in which the interaction and the resulting effective phantom behaviour follow directly from the underlying field-space geometry and symmetry-breaking structure.

%%% 5. Dependence on parameters and ICs
\vspace{0.3cm}

\paragraph*{\textbf{Dependence on parameters and initial conditions}}

We now briefly discuss the dependence of the cosmological evolution on the model parameters and initial conditions. The corresponding numerical results are presented in Appendix~\ref{Appendix}, where each quantity is varied individually. We do not fit them to give the correct abundances, however. 

The importance of the interaction at early times is controlled primarily by $m_\varphi$ and $\varphi_i/f_\theta$. Close to $\theta/f_\theta=\pi/2$, the curvature of the effective potential is approximately $m_{\theta,{\rm eff}}^2\simeq2m_\theta^2-\frac{m_\varphi^2}{4}\left[1-\cos(4\varphi_i/f_\theta)\right]$. Increasing either $m_\varphi$ or $\varphi_i/f_\theta$ strengthens the interaction relative to the self-potential. If this combination is too small, the interaction has little effect on the evolution of $\theta$, and the cosmology approaches the weakly interacting limit of scalar-field DM and quintessence.

The mass $m_\varphi$ also determines when the DM field begins to oscillate, through the condition $m_\varphi|\sin(\theta/f_\theta)|\sim H$. In the underlying theory, $m_\varphi$ is set by the explicit $O(3)$-breaking parameter $\epsilon_2$, with $m_\varphi^2=2\epsilon_2M_p^2/\xi=2\epsilon_2f_\theta^2$. Its value is therefore not fixed by the construction, and the DM field can, in principle, lie in a different mass range. In this work, we focus on the ultralight regime, $m_\varphi\sim10^{-20}\,\mathrm{eV}$. A larger $m_\varphi$ leads to an earlier onset of the DM oscillations and an earlier decay of their amplitude. By contrast, $\varphi_i/f_\theta$ mainly controls the initial DM abundance and the magnitude of its contribution to the interaction. These two quantities are therefore correlated once the present-day DM abundance is imposed.

The DE mass $m_\theta$ controls the curvature of the self-potential and the timescale of the late-time evolution of $\theta$. In the underlying theory, $m_\theta$ is set by the explicit $O(3)$-breaking parameter $\epsilon_1$, with $m_\theta^2=\epsilon_1M_p^2/\xi=\epsilon_1f_\theta^2$. Its value is therefore not fixed by the construction. Increasing $m_\theta$ causes the field to respond to its self-potential and begin its late-time motion earlier, thereby shifting the redshift and duration of the dynamical and effective phantom phases. To retain DE-like behaviour until recent times, one requires $m_\theta\lesssim H_0$. As discussed above, $m_\theta$ controls the dynamics of the DE field, but does not by itself determine the epoch at which its energy density becomes dominant.

The initial value $\theta_i/f_\theta$ determines the displacement of the DE field from the unstable point at $\theta/f_\theta=\pi/2$. This displacement seeds the interaction-driven motion of $\theta$: if the field were placed exactly at the maximum with vanishing velocity, it would remain there at the homogeneous classical level. A larger initial displacement generally produces a stronger early-time response and a larger transient kinetic contribution. At the same time, $\theta_i/f_\theta$ contributes to the normalization of the DE energy density and is consequently constrained by the requirement of reproducing its present-day abundance. Initial values that are too far from $\pi/2$ may also drive the evolution towards the boundary of the coordinate chart before the self-potential becomes dominant.

Finally, $\xi$ determines the field-space scale through $f_\theta=M_p/\sqrt{\xi}$. At fixed values of the dimensionless field ratios, changing $\xi$ changes the absolute field range and the normalization of the scalar energy densities. It does not independently determine whether the interaction dominates over the self-potential, since both contributions contain the same overall factor of $f_\theta^2$. Once the present-day abundances are imposed, variations of $\xi$ must therefore be accompanied by corresponding changes in the masses or initial field values.

The model is thus not equally sensitive to all five quantities. The existence of an early interaction-dominated phase is primarily controlled by the combination of $m_\varphi$, $m_\theta$, and $\varphi_i/f_\theta$, while $\theta_i/f_\theta$ determines how the instability is initially seeded. The late-time DE dynamics is governed mainly by $m_\theta$, and $\xi$ fixes the overall field-space. 
The results presented in Appendix~\ref{Appendix} illustrate the degeneracy between the parameters, and the particular effect of each one. Note that we do not fit the parameters in the appendix to give the correct abundances. As can be seen from the plots, there is a degeneracy among the parameters, such that the fitting of the initial conditions would partially obscure the effect of changing the masses. This indicates the possibility that the model may hold across a larger range of parameter space, although we do not perform a systematic parameter scan. A complete determination of the viable parameter space, including observational constraints, would require a dedicated joint scan of all parameters and initial conditions.

\section{Conclusion}\label{sec:conclusion}

In this work, we have constructed an interacting dark-sector model within an $O(3)$ extension of No-Scale Gravity. In the Einstein frame, the radial degree of freedom is the dilaton, while the two angular directions are identified with DE and DM. Their masses and interaction arise from the same symmetry-breaking sector. The dark components and their coupling are therefore not introduced independently, but emerge from a common gravitational framework.

We have shown that, at the background level, the model reproduces the required cosmological evolution. The heavier field $\varphi$ undergoes coherent oscillations and behaves as pressureless scalar-field DM, while the lighter field $\theta$ remains nearly frozen throughout most of the cosmological history and becomes dynamical at late times, driving accelerated expansion. The numerical solutions reproduce the standard cosmological evolution, together with the required present-day abundances. 

The interaction produces a characteristic evolution of the effective potential of $\theta$. At early times, the DM field destabilizes the initial position of $\theta$ and induces a transient period of kinetic domination, during which $w_\theta\simeq1$. This component remains extremely subdominant and does not affect the background expansion. As the oscillation amplitude of $\varphi$ decreases, the interaction is suppressed, and the self-potential of $\theta$ becomes dominant, leading to its late-time dynamical DE evolution.

Although the fundamental DE field is canonical and always satisfies $w_\theta\geq-1$, the interacting dark sector can be interpreted as having an effective EoS that crosses below $-1$. The apparent phantom behaviour emerges when the interacting cosmology is interpreted under the assumption of non-interacting DM with the standard $a^{-3}$ scaling. For the region explored here, the crossing occurs at redshifts $0.5\lesssim z\lesssim2.5$, producing the type of effective phantom evolution suggested by combinations of DESI and other cosmological observations, without introducing a fundamental phantom degree of freedom.

Beyond this late-time DE phenomenology, the construction also allows for a broader range of DM realizations. The DM mass is determined by the symmetry-breaking parameter and is not fixed to the ultralight scale considered here. Heavier scalar-DM realizations of the same construction are therefore possible. The ultralight regime is nevertheless particularly interesting, since it opens the possibility of exploring the characteristic wave-like astrophysical and cosmological phenomenology of ultralight DM within an interacting and gravitationally unified dark sector~\cite{Marsh:2015xka,Ferreira:2020fam,Hui:2021tkt,Eberhardt:2025caq}.

The construction provides a fundamental realization of interacting DE within No-Scale Gravity, in which DM, DE, and their interaction arise from a common underlying symmetry structure. It gives a theoretical foundation to interacting dark-sector scenarios that have proved phenomenologically successful in the light of DESI. The next step is to develop the cosmological perturbations of the model and perform a direct comparison with CMB, BAO, supernova, and other large-scale-structure data, in order to determine its observationally viable parameter space and test its effective phantom evolution.

%%%%%%%%%%%%%%%%%%%%%%%%%%%%%%%%%%%
\textbf{\textit{Acknowledgements}}
%%%%%%%%%%%%%%%%%%%%%%%%%%%%%%%%%%%

 M. H. thanks Derek Inman for discussion. This work was supported in part by JSPS KAKENHI No. 25K17381 (E.~G.~M.~F.), JP24H02244 (T.~T.~Y.) and E.~G.~M.~F. were also supported by World Premier International Research Center Initiative (WPI Initiative), MEXT, Japan. L.~dS.~P. was supported by “Fundação de Amparo à Pesquisa do Estado de São Paulo” (FAPESP) under contracts 2025/24242-0 and 2024/16149-8.

\bibliographystyle{unsrt}
\bibliography{references}

@article{Hong:2025cae,
    author = "Hong, Muzi and Mukaida, Kyohei and Yanagida, Tsutomu T.",
    title = "{Dynamical dark energy in the no-scale Brans{\textendash}Dicke gravity}",
    eprint = "2506.01543",
    archivePrefix = "arXiv",
    primaryClass = "hep-ph",
    doi = "10.1016/j.physletb.2025.139932",
    journal = "Phys. Lett. B",
    volume = "870",
    pages = "139932",
    year = "2025"
}

@article{Hong:2025tyi,
    author = "Hong, Muzi and Mukaida, Kyohei and Yanagida, Tsutomu T.",
    title = "{No-scale Brans-Dicke Gravity -- ultralight scalar boson {\&} heavy inflaton}",
    eprint = "2503.18648",
    archivePrefix = "arXiv",
    primaryClass = "hep-ph",
    doi = "10.1088/1475-7516/2025/10/032",
    journal = "JCAP",
    volume = "10",
    pages = "032",
    year = "2025"
}

@article{Lin:2025gne,
    author = "Lin, Weikang and Visinelli, Luca and Yanagida, Tsutomu T.",
    title = "{Testing quintessence axion dark energy with recent cosmological results}",
    eprint = "2504.17638",
    archivePrefix = "arXiv",
    primaryClass = "astro-ph.CO",
    doi = "10.1088/1475-7516/2025/10/023",
    journal = "JCAP",
    volume = "10",
    pages = "023",
    year = "2025"
}

@article{Shaposhnikov:2008xi,
    author = "Shaposhnikov, Mikhail and Zenhausern, Daniel",
    title = "{Quantum scale invariance, cosmological constant and hierarchy problem}",
    eprint = "0809.3406",
    archivePrefix = "arXiv",
    primaryClass = "hep-th",
    doi = "10.1016/j.physletb.2008.11.041",
    journal = "Phys. Lett. B",
    volume = "671",
    pages = "162--166",
    year = "2009"
}

@article{Armillis:2013wya,
    author = "Armillis, Roberta and Monin, Alexander and Shaposhnikov, Mikhail",
    title = "{Spontaneously Broken Conformal Symmetry: Dealing with the Trace Anomaly}",
    eprint = "1302.5619",
    archivePrefix = "arXiv",
    primaryClass = "hep-th",
    doi = "10.1007/JHEP10(2013)030",
    journal = "JHEP",
    volume = "10",
    pages = "030",
    year = "2013"
}

@article{Dutta:2010cu,
    author = "Dutta, Sourish and Scherrer, Robert J.",
    title = "{Big Bang nucleosynthesis with a stiff fluid}",
    eprint = "1006.4166",
    archivePrefix = "arXiv",
    primaryClass = "astro-ph.CO",
    doi = "10.1103/PhysRevD.82.083501",
    journal = "Phys. Rev. D",
    volume = "82",
    pages = "083501",
    year = "2010"
}

@article{Das:2005yj,
    author = "Das, Subinoy and Corasaniti, Pier Stefano and Khoury, Justin",
    title = "{Super-acceleration as signature of dark sector interaction}",
    eprint = "astro-ph/0510628",
    archivePrefix = "arXiv",
    doi = "10.1103/PhysRevD.73.083509",
    journal = "Phys. Rev. D",
    volume = "73",
    pages = "083509",
    year = "2006"
}

@article{Wetterich:2019qzx,
    author = "Wetterich, C.",
    title = "{Quantum scale symmetry}",
    eprint = "1901.04741",
    archivePrefix = "arXiv",
    primaryClass = "hep-th",
    month = "1",
    year = "2019"
}

@article{Brans:1961sx,
    author = "Brans, C. and Dicke, R. H.",
    editor = "Hsu, Jong-Ping and Fine, D.",
    title = "{Mach's principle and a relativistic theory of gravitation}",
    doi = "10.1103/PhysRev.124.925",
    journal = "Phys. Rev.",
    volume = "124",
    pages = "925--935",
    year = "1961"
}

@article{Kawasaki:2023zpd,
    author = "Kawasaki, Masahiro and Yanagida, Tsutomu T.",
    title = "{Hill-top inflation from Dai-Freed anomaly in the standard model {\textemdash} a solution to the iso-curvature problem of the axion dark matter}",
    eprint = "2306.14579",
    archivePrefix = "arXiv",
    primaryClass = "hep-ph",
    doi = "10.1088/1475-7516/2024/01/014",
    journal = "JCAP",
    volume = "01",
    pages = "014",
    year = "2024"
}

@article{Englert:1976ep,
    author = "Englert, F. and Truffin, C. and Gastmans, R.",
    title = "{Conformal Invariance in Quantum Gravity}",
    reportNumber = "PRINT-76-0296 (BRUSSELS)",
    doi = "10.1016/0550-3213(76)90406-5",
    journal = "Nucl. Phys. B",
    volume = "117",
    pages = "407--432",
    year = "1976"
}

@article{Hamada:2016onh,
    author = "Hamada, Yuta and Kawai, Hikaru and Nakanishi, Yukari and Oda, Kin-ya",
    title = "{Meaning of the field dependence of the renormalization scale in Higgs inflation}",
    eprint = "1610.05885",
    archivePrefix = "arXiv",
    primaryClass = "hep-th",
    reportNumber = "OU-HET-906, KEK-TH-1926, MAD-TH-16-09",
    doi = "10.1103/PhysRevD.95.103524",
    journal = "Phys. Rev. D",
    volume = "95",
    number = "10",
    pages = "103524",
    year = "2017"
}

@article{Falls:2018olk,
    author = "Falls, Kevin and Herrero-Valea, Mario",
    title = "{Frame (In)equivalence in Quantum Field Theory and Cosmology}",
    eprint = "1812.08187",
    archivePrefix = "arXiv",
    primaryClass = "hep-th",
    doi = "10.1140/epjc/s10052-019-7070-3",
    journal = "Eur. Phys. J. C",
    volume = "79",
    number = "7",
    pages = "595",
    year = "2019"
}

@article{Will:2005va,
    author = "Will, Clifford M.",
    title = "{The Confrontation between general relativity and experiment}",
    eprint = "gr-qc/0510072",
    archivePrefix = "arXiv",
    doi = "10.12942/lrr-2006-3",
    journal = "Living Rev. Rel.",
    volume = "9",
    pages = "3",
    year = "2006"
}

@article{Ferreira:2016kxi,
    author = "Ferreira, Pedro G. and Hill, Christopher T. and Ross, Graham G.",
    title = "{No fifth force in a scale invariant universe}",
    eprint = "1612.03157",
    archivePrefix = "arXiv",
    primaryClass = "gr-qc",
    reportNumber = "FERMILAB-PUB-16-665-T",
    doi = "10.1103/PhysRevD.95.064038",
    journal = "Phys. Rev. D",
    volume = "95",
    number = "6",
    pages = "064038",
    year = "2017"
}

@article{Burrage:2018dvt,
    author = "Burrage, Clare and Copeland, Edmund J. and Millington, Peter and Spannowsky, Michael",
    title = "{Fifth forces, Higgs portals and broken scale invariance}",
    eprint = "1804.07180",
    archivePrefix = "arXiv",
    primaryClass = "hep-th",
    reportNumber = "IPPP/18/23, IPPP-18-23",
    doi = "10.1088/1475-7516/2018/11/036",
    journal = "JCAP",
    volume = "11",
    pages = "036",
    year = "2018"
}

@article{DESI:2013agm,
    author = "Levi, Michael and others",
    collaboration = "DESI",
    title = "{The DESI Experiment, a whitepaper for Snowmass 2013}",
    eprint = "1308.0847",
    archivePrefix = "arXiv",
    primaryClass = "astro-ph.CO",
    month = "8",
    year = "2013"
}

@article{DESI:2016fyo,
    author = "Aghamousa, Amir and others",
    collaboration = "DESI",
    title = "{The DESI Experiment Part I: Science,Targeting, and Survey Design}",
    eprint = "1611.00036",
    archivePrefix = "arXiv",
    primaryClass = "astro-ph.IM",
    reportNumber = "FERMILAB-PUB-16-517-AE",
    month = "10",
    year = "2016"
}

@article{DESI:2016igz,
    author = "Aghamousa, Amir and others",
    collaboration = "DESI",
    title = "{The DESI Experiment Part II: Instrument Design}",
    eprint = "1611.00037",
    archivePrefix = "arXiv",
    primaryClass = "astro-ph.IM",
    reportNumber = "FERMILAB-PUB-16-518-AE",
    month = "10",
    year = "2016"
}

@article{DESI:2025zpo,
    author = "Abdul Karim, M. and others",
    collaboration = "DESI",
    title = "{DESI DR2 Results I: Baryon Acoustic Oscillations from the Lyman Alpha Forest}",
    eprint = "2503.14739",
    archivePrefix = "arXiv",
    primaryClass = "astro-ph.CO",
    reportNumber = "FERMILAB-PUB-25-0167-PPD",
    month = "3",
    year = "2025"
}

@article{DESI:2025zgx,
    author = "Abdul Karim, M. and others",
    collaboration = "DESI",
    title = "{DESI DR2 Results II: Measurements of Baryon Acoustic Oscillations and Cosmological Constraints}",
    eprint = "2503.14738",
    archivePrefix = "arXiv",
    primaryClass = "astro-ph.CO",
    reportNumber = "FERMILAB-PUB-25-0169-PPD",
    month = "3",
    year = "2025"
}

@article{Eberhardt:2025caq,
    author = "Eberhardt, Andrew and Ferreira, Elisa G. M.",
    title = "{Ultralight fuzzy dark matter review}",
    eprint = "2507.00705",
    archivePrefix = "arXiv",
    primaryClass = "astro-ph.CO",
    month = "7",
    year = "2025"
}

@article{Ferreira:2020fam,
    author = "Ferreira, Elisa G. M.",
    title = "{Ultra-light dark matter}",
    eprint = "2005.03254",
    archivePrefix = "arXiv",
    primaryClass = "astro-ph.CO",
    doi = "10.1007/s00159-021-00135-6",
    journal = "Astron. Astrophys. Rev.",
    volume = "29",
    number = "1",
    pages = "7",
    year = "2021"
}

@article{Marsh:2015xka,
    author = "Marsh, David J. E.",
    title = "{Axion Cosmology}",
    eprint = "1510.07633",
    archivePrefix = "arXiv",
    primaryClass = "astro-ph.CO",
    reportNumber = "KCL-PH-TH-2015-50",
    doi = "10.1016/j.physrep.2016.06.005",
    journal = "Phys. Rept.",
    volume = "643",
    pages = "1--79",
    year = "2016"
}

@article{Hui:2021tkt,
    author = "Hui, Lam",
    title = "{Wave Dark Matter}",
    eprint = "2101.11735",
    archivePrefix = "arXiv",
    primaryClass = "astro-ph.CO",
    doi = "10.1146/annurev-astro-120920-010024",
    journal = "Ann. Rev. Astron. Astrophys.",
    volume = "59",
    pages = "247--289",
    year = "2021"
}

@article{Poulin:2018dzj,
    author = "Poulin, Vivian and Smith, Tristan L. and Grin, Daniel and Karwal, Tanvi and Kamionkowski, Marc",
    title = "{Cosmological implications of ultralight axionlike fields}",
    eprint = "1806.10608",
    archivePrefix = "arXiv",
    primaryClass = "astro-ph.CO",
    doi = "10.1103/PhysRevD.98.083525",
    journal = "Phys. Rev. D",
    volume = "98",
    number = "8",
    pages = "083525",
    year = "2018"
}

@article{Hlozek:2014lca,
    author = "Hlozek, Ren{\'e}e and Grin, Daniel and Marsh, David J. E. and Ferreira, Pedro G.",
    title = "{A search for ultralight axions using precision cosmological data}",
    eprint = "1410.2896",
    archivePrefix = "arXiv",
    primaryClass = "astro-ph.CO",
    doi = "10.1103/PhysRevD.91.103512",
    journal = "Phys. Rev. D",
    volume = "91",
    number = "10",
    pages = "103512",
    year = "2015"
}

@article{Giare:2024smz,
    author = "Giar{\`e}, William and Sabogal, Miguel A. and Nunes, Rafael C. and Di Valentino, Eleonora",
    title = "{Interacting Dark Energy after DESI Baryon Acoustic Oscillation Measurements}",
    eprint = "2404.15232",
    archivePrefix = "arXiv",
    primaryClass = "astro-ph.CO",
    doi = "10.1103/PhysRevLett.133.251003",
    journal = "Phys. Rev. Lett.",
    volume = "133",
    number = "25",
    pages = "251003",
    year = "2024"
}

@article{Antusch:2026ldp,
    author = "Antusch, Stefan and King, Stephen F. and Wang, Xin",
    title = "{Coupled Dark Energy and Dark Matter for DESI: An Effective Guide to the Phantom Divide}",
    eprint = "2604.08449",
    archivePrefix = "arXiv",
    primaryClass = "astro-ph.CO",
    month = "4",
    year = "2026"
}

@article{Guedezounme:2025wav,
    author = "Guedezounme, S{\^e}cloka L. and Dinda, Bikash R. and Maartens, Roy",
    title = "{Phantom crossing or dark interaction?}",
    eprint = "2507.18274",
    archivePrefix = "arXiv",
    primaryClass = "astro-ph.CO",
    doi = "10.1088/1475-7516/2026/01/062",
    journal = "JCAP",
    volume = "01",
    pages = "062",
    year = "2026"
}

@article{Planck:2018vyg,
    author = "Aghanim, N. and others",
    collaboration = "Planck",
    title = "{Planck 2018 results. VI. Cosmological parameters}",
    eprint = "1807.06209",
    archivePrefix = "arXiv",
    primaryClass = "astro-ph.CO",
    doi = "10.1051/0004-6361/201833910",
    journal = "Astron. Astrophys.",
    volume = "641",
    pages = "A6",
    year = "2020",
    note = "[Erratum: Astron.Astrophys. 652, C4 (2021)]"
}

\clearpage
\appendix
\setcounter{equation}{0}
\setcounter{table}{0}
\setcounter{figure}{0}
\renewcommand{\thetable}{A\Roman{table}}
\renewcommand{\thefigure}{A\arabic{figure}}
\renewcommand{\theequation}{A\arabic{equation}}

%%%%%%%%%%%%%%%%%%%%%%%%%%%%%%%%%%%
%%%%%%%%%%%%%%%%%%%%%%%%%%%%%%%%%%%
%%%%%%%%%%%%%%%%%%%%%%%%%%%%%%%%%%%
%%%%%%%%%%%%%%%%%%%%%%%%%%%%%%%%%%%
% \section{Appendix}\label{app:}

%%%%%%%%%%%%%%%%%%%%%%%%%%%%%%%%%%%
%%%%%%%%%%%%%%%%%%%%%%%%%%%%%%%%%%%
%%%%%%%%%%%%%%%%%%%%%%%%%%%%%%%%%%%
%%%%%%%%%%%%%%%%%%%%%%%%%%%%%%%%%%%
\section{Parameter sensitivity}\label{Appendix}

To illustrate the robustness of the cosmological evolution and the role of each model parameter, we present the field evolution, energy densities, and equations of state for variations of $\{\xi,\varphi_i/f_\theta,\theta_i/f_\theta,m_\theta,m_\varphi\}$. We use the baseline values $\xi=10^{-4}$, $\varphi_i/f_\theta=10^{-4}$, $\theta_i/f_\theta=\pi/2-10^{-1}$, $m_\theta=1.4\times10^{-1}H_0$, and $m_\varphi=2\times10^{13}H_0$, and vary one quantity at a time, while keeping the remaining variables fixed. These results illustrate the qualitative parameter dependence discussed in Section~\ref{sec:Numerics}. We do not, however, fit the correct abundances in this appendix, so as not to obscure the effects of changing the masses and the initial conditions. 

\begin{figure}[!htbp]
    \centering
    \includegraphics[width=0.699\textwidth]{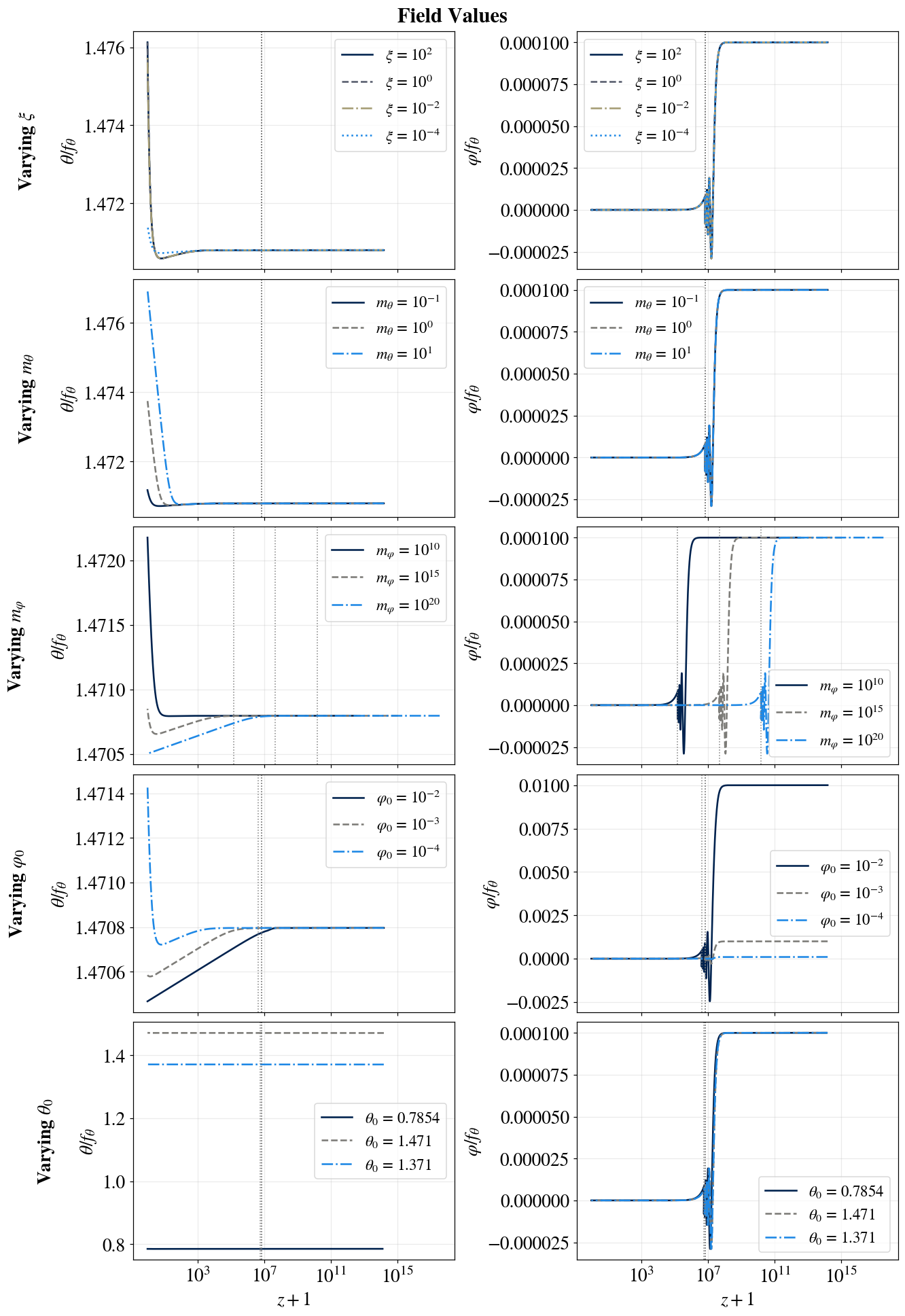}
    \caption{Plots of field value evolution as a function of redshift. Left and right panels show $\theta/f_\theta$ and $\varphi/f_\theta$, respectively. All plots share the baseline parameters: $\varphi_i/f_\theta = 10^{-4}$, $\theta_i/f_\theta = \pi/2 - 10^{-1}$, $m_\theta = 1.4 \times 10^{-1} \, H_0$, $m_\varphi = 2 \times 10^{13}\, H_0$, and $\xi = 10^{-4}$. Rows from top to bottom vary: (1) $\xi$, (2) $m_\theta$, (3) $m_\varphi$, (4) $\theta_i/f_\theta$, and (5) $\varphi_i/f_\theta$.}
    \label{fig:appendix_fields}
\end{figure}

\begin{figure}[htbp]
    \centering
    \includegraphics[width=0.73\textwidth]{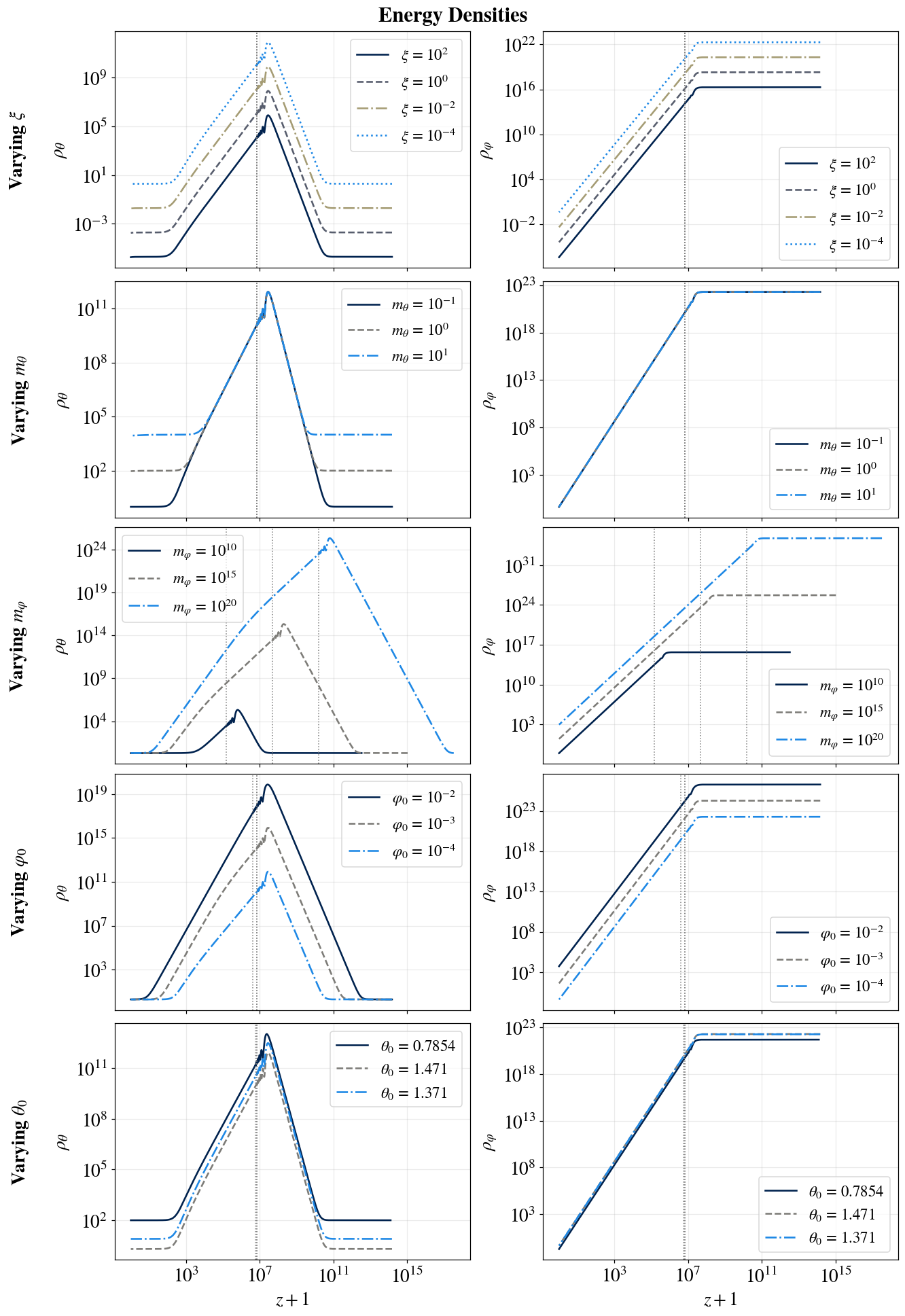}
    \caption{Plots of energy density evolution as a function of redshift. Left and right panels show $\rho_\theta$ and $\rho_\varphi$, respectively. All plots share the baseline parameters: $\varphi_i/f_\theta = 10^{-4}$, $\theta_i/f_\theta = \pi/2 - 10^{-1}$, $m_\theta = 1.4 \times 10^{-1} \, H_0$, $m_\varphi = 2 \times 10^{13}\, H_0$, and $\xi = 10^{-4}$. Rows from top to bottom vary: (1) $\xi$, (2) $m_\theta$, (3) $m_\varphi$, (4) $\theta_i/f_\theta$, and (5) $\varphi_i/f_\theta$.}
    \label{fig:appendix_densities}
\end{figure}

\begin{figure}[htbp]
    \centering
    \includegraphics[width=0.73\textwidth]{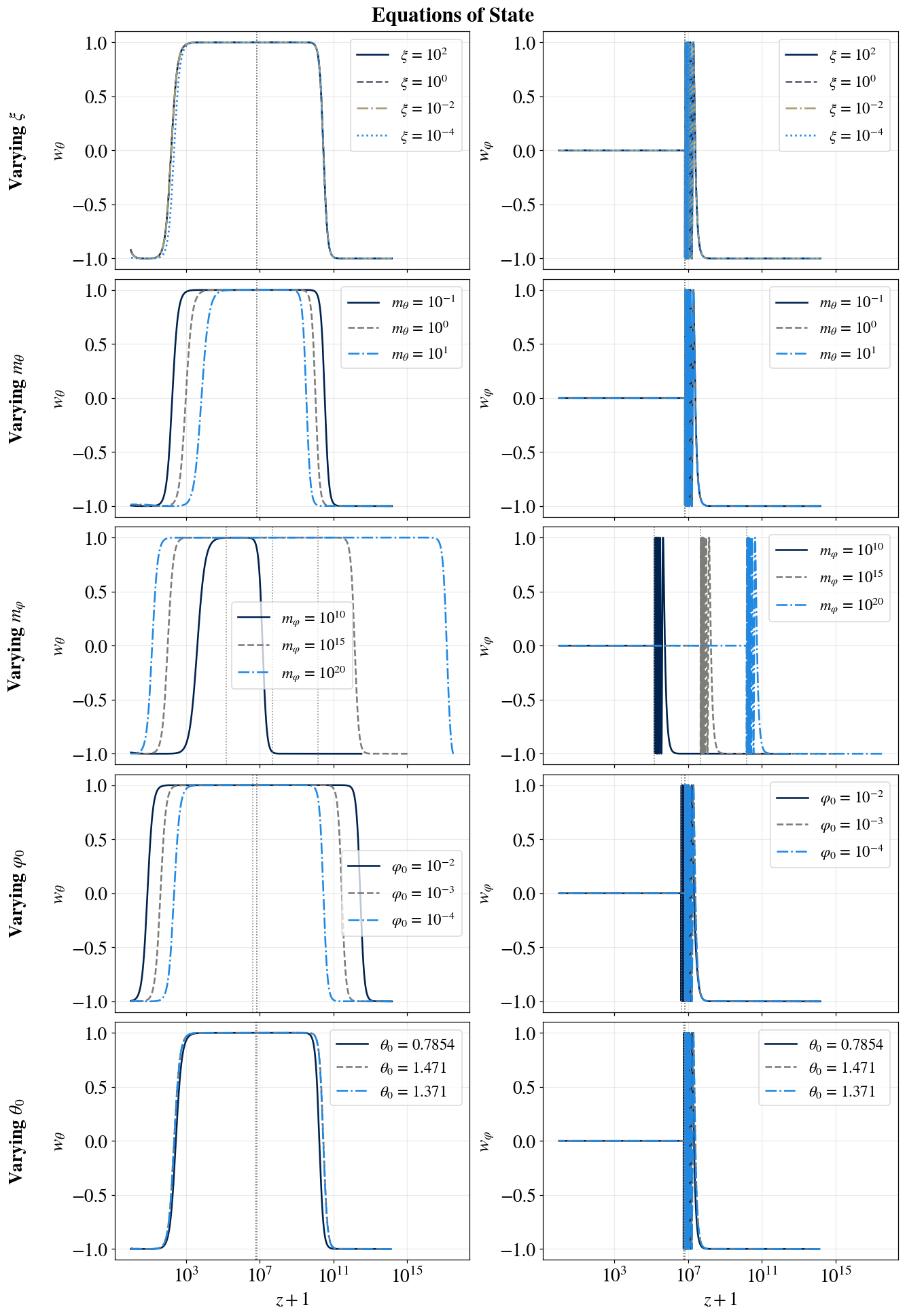}
    \caption{Plots of EoS evolution as a function of redshift. Left and right panels show $w_\theta$ and $w_\varphi$, respectively. All plots share the baseline parameters: $\varphi_i/f_\theta = 10^{-4}$, $\theta_i/f_\theta = \pi/2 - 10^{-1}$, $m_\theta = 1.4 \times 10^{-1} \, H_0$, $m_\varphi = 2 \times 10^{13} \, H_0$, and $\xi = 10^{-4}$. Rows from top to bottom vary: (1) $\xi$, (2) $m_\theta$, (3) $m_\varphi$, (4) $\theta_i/f_\theta$, and (5) $\varphi_i/f_\theta$.}
    \label{fig:appendix_eos}
\end{figure}

\end{document}